\documentclass[review]{elsarticle}
\usepackage{csquotes}
\usepackage{array}
\usepackage{amsmath}
\usepackage{graphicx}
\usepackage{subcaption}
\usepackage{algpseudocode}
\usepackage{cleveref}
\usepackage{lipsum}
\usepackage{natbib}
\usepackage{stfloats}
\usepackage{mathcomp}
\usepackage{textcomp}
\usepackage{float}
\usepackage{gensymb}
\usepackage{csquotes}
\usepackage{graphics}
\usepackage{epstopdf}
\usepackage{eqnarray}
\usepackage{multirow}
\usepackage{amssymb}
\usepackage{url}
\usepackage{xcolor}
\usepackage{hhline}
\usepackage[linesnumbered,ruled,vlined]{algorithm2e}
\newcolumntype{P}[1]{>{\centering\arraybackslash}p{#1}}
\usepackage[T1]{fontenc}
\usepackage{enumitem}
\usepackage{ctable}
\usepackage{pifont}
\usepackage{caption}
\usepackage{dsfont}
\usepackage{csquotes}
\usepackage{graphics}
\usepackage{epstopdf}
\usepackage{color,soul}
\usepackage[T1]{fontenc}
\usepackage{tfrupee}
\let\orupee\rupee
\def\rupee{\ifmmode\text{\orupee}\else\orupee\fi}
\usepackage{url}

\usepackage{amsmath}

\usepackage{nomencl}
\emergencystretch=\maxdimen
\begin{document}

\title{A Multi-Head Convolutional Neural Network Based Non-Intrusive Load Monitoring Algorithm Under Dynamic Grid Voltage Conditions}

\author{Himanshu Grover, Lokesh Panwar, Ashu Verma, B. K. Panigrahi$^*$,T. S. Bhatti}
\address{Department of Energy Science and Engineering, Indian Institute of Technology Delhi\\ $^*$ Department of Electrical Engineering, Indian Institute of Technology Delhi
\vspace{-1.2cm}
}
\begin{abstract}
In recent times, non-intrusive load monitoring (NILM) has emerged as an important tool for distribution-level energy management systems owing to its potential for energy conservation and management. However, load monitoring in smart building environments is challenging due to high variability of real-time load and varied load composition. Furthermore, as the volume and dimensionality of smart meter's data increases, accuracy and computational time are key concerning factors. In view of these challenges, this paper proposes an improved NILM technique using multi-head (Mh-Net) convolutional neural network (CNN) under dynamic grid voltage conditions. An attention layer is introduced into  the  proposed CNN model, which  helps in improving estimation accuracy of appliance power consumption. The performance of the developed model has been verified on an experimental laboratory setup for multiple appliance sets with varied power consumption levels, under dynamic grid voltages. Moreover, the effectiveness of the proposed model has been verified on widely used UK-DALE data, and its performance has been compared with existing NILM techniques. Results depict that the proposed model accurately identifies appliances, power consumptions and their time-of-use even during practical dynamic grid voltage conditions.
\end{abstract}
\begin{keyword}
Non-intrusive load monitoring (NILM), Machine learning (ML), Convolution Neural Network (CNN), Multi-head CNN (MhNet), Attention Layer, Load disaggregation.
\end{keyword}
\maketitle
\section{Introduction}
\indent Modernization of power systems is associated with the emergence of smart grids, along with widespread penetration of distributed energy resources (DERs) and increased incorporation of information and communication technology (ICT). The building blocks of a smart grid are constituted by smart buildings, that enable interaction between the utility and consumers in multiple ways, including real-time energy monitoring, load monitoring and demand-side management techniques etc.
Recent reports and statistics reveal that buildings have emerged to be major electricity consumers, and their energy consumption is expected to gradually increase by 1.5\% annually from 2012 to 2040 \cite{9428578}. Despite being prominent energy consumers, buildings have enormous potential of energy conservation and management due to the wide range of user-appliances and high operational flexibility.
To harness the energy conservation potential of smart buildings, it is imperative to understand the appliance behaviour, utilization frequency, their time-of-use (ToU) and energy consumption \cite{9078031}. Smart metering infrastructure plays a vital role in acquiring such information, 
providing benefits to both the utility and consumers in real-time load monitoring, analysis of energy consumption, energy optimization and control.\\
\indent Among the various energy conservation approaches enabled by smart metering, an important tool in context of smart buildings and demand side management is load monitoring.  Load  monitoring  helps  the  utility  as  well  as  users in  providing  information  such as energy  audit, demand response (DR), power flow analysis, energy  management and power security etc. \cite{8745491,9320585}. Strategies for load monitoring are broadly classified into intrusive load monitoring (ILM) and non-intrusive load monitoring (NILM) \cite{8411449}. An ILM technique utilizes sensors for each appliance which accurately monitor its power consumption and ToU. Since, ILM requires sensors for each appliance, the implementation cost and complexity of such a load monitoring technique is quite high \cite{8316882}. In contrast NILM is the process of obtaining information about energy consumption, measured and aggregated by the end-users through smart meters \cite{9208737}. Unlike ILM, NILM technique does not utilize individual sensors for each appliance. However, NILM utilizes single point aggregated data from a smart meter which measures overall building energy consumption, thus reducing the overall installation cost. Motivated by these benefits, NILM has become an important tool for the utility, as it plays a vital role in energy conservation and enabling DR programs. This has been further assisted by recent advancements in distribution system infrastructure through the integration of smart meters and integration control technologies.\\ 
\indent Based on the learning methodologies, NILM techniques are broadly classified as unsupervised, semi-supervised and supervised NILM. Unsupervised NILM utilizes the aggregated data without any prior information about the appliance power consumption. Based on two auxiliary algorithms, the work in \cite{8745491} proposes a method for acquiring appliance energy consumption for relatively complex household  datasets. Motivated by the operation of Hidden Markov Model with basic functions for modelling, it is widely adopted by many researchers \cite{PARSON20141,7605545,7750555}.  
However, since unsupervised NILM does not utilize past data for its model training, its performance may severly deteriorate under practical operating conditions. This may result from the fact that since appliance load profiles vary with dynamic grid conditions, unsupervised NILM models may not be able to accurately identify the appliances. Moreover, variation in sampling frequency may lead to feature loss, further reducing the accuracy. 

Semi-supervised NILM adopts multiple sensors for group of appliances and are mostly applied for classification of appliances \cite{9380223}. Instead of utilizing single-point aggregated data for event detection, it utilizes several aggregated datapoints for acquiring information about individual appliances \cite{7210212}. Yang et \textit{al.} in \cite{8911216} developed a semi-supervised NILM technique for a multilabel classification problem which occurs mainly due to simultaneous operation of appliances. In \cite{9119146}, hall effect sensors were installed on wires connecting appliances, which reduces the estimation error. However, the installation of additional sensors increases the solution cost. Consequently, the accuracy of supervised NILM is quite high as compared to unsupervised and semi-supervised learning approaches, supervised NILM eliminates the need for individual sensors for the appliances, as it requires only datasets for training the classifier which helps in identification of appliance operation from the aggregated smart meter data.\\
\indent Supervised NILM are trained with individual appliance datasets and appropriately extract information from aggregated energy meter data. A deep learning approach for NILM is proposed in \cite{7847445} using multiple layers of dictionaries. Similarly, deep dictionary and deep transform learning approach is proposed in \cite{8315456} for simultaneous detection of multiple appliances. Further, authors in \cite{8720065} developed an NILM technique using deep learning and post processing for Type-II appliances. Literature studies reveals that even though numerous papers utilized supervised NILM models, most existing works either focus on \enquote{only low} or \enquote{only high} power consumption appliances which further deteriorates the operational accuracy of the models under practical scenarios. Thus, the requirement of high accuracy is still a major concern which can be improved using machine learning (ML) methods like CNN, recurrent neural network (RNN), denoising auto-encoders (DAE) \cite{Kelly2015NeuralND}\cite{He2016/11}.\\
\indent A close review of existing literature reveals that CNN demonstrates good performance in NILM applications. Owing to its flexibility in application and modeling, various CNN structures have been proposed such as sequence-to-point and sequence-to-sequence models \cite{bc925f015ff2472d8af9ec4886b91b9e}. The application of CNN has also created a new direction of multi-scale information extraction, which is usually used in computer vision or applications \cite{8720065}. There are many ways to associate multi-scale features such as different sampling in different layers and structural changes in the CNN \cite{Bahdanau2015NeuralMT,Sutskever2014SequenceTS,article3}. Another approach to associate multi scale information into the model uses dilated convolutions \cite{young2017recent,Raffel2015FeedForwardNW}, which is simple to adopt without requiring structural changes in the model. 
However, it must be noted that while CNN application is faced with limitations in processing long-distance information, recurrent neural
networks (RNN) application is faced with difficulty in training due to their sequential characteristics. In view of this, multi-head(Mh-Net) CNN demonstrates superior performance in processing time-series data, enabling the networks to identify and learn multi-level dependencies in data. This further facilitates parallel training of the network, thereby saving processing time and costs \cite{9259533}.\\
\indent It is further noted that traditionally encoders and decoders are used to extract and understand the context of information \cite{Sutskever2014SequenceTS}. However, such an architecture is adopted for small input information only because with large information, the output tends to be distorted. Furthermore, expansion of the networks for large-scale problems may result in over-fitting and convergence problems \cite{KHAN2021107671}. In view of these challenges, recent research works have introduced the attention layer which demonstrates superior performance to understand large information, yielding better classification outputs through vital feature selection \cite{en12112203}. Thus, the attention layer yields improved and optimal performance, resulting in automatic extraction of important features.\\
\indent A crucial observation of existing NILM models indicates that training of existing techniques is carried out under constant grid voltage conditions. However, the hypothesis does not hold for practical scenarios since due to the dynamic nature of grid voltage, appliance power profiles may observe variations under different grid voltage conditions. Under practical conditions, the training dataset (at constant voltage) and target dataset (at dynamic grid voltage) may not  be identical, which may degrade the performance when the NILM model trained with training dataset encounters the practical dataset under dynamic grid voltage conditions. To the best of the authors' knowledge, the impact of NILM performance under such dynamic grid voltage scenario has not been explored much in existing literature. \\
\indent In view of the aforementioned discussion, this paper focuses on development of a Mh-Net CNN for load monitoring in smart buildings through precise identification of appliances and their operational schedules during real time operation. In order to improve the accuracy of load monitoring while processing large volume datasets an attention layer has been included in the proposed Mh-Net CNN model. To further improve the performance of the developed NILM model for DR and grid-support applications under practical operating conditions, the proposed model has been trained under dynamic grid voltage conditions. 
The major contributions of this work are summarized as follows.
\begin{enumerate}
    \item A supervised NILM technique using Mh-Net CNN model has been developed for identification of appliance power and ToU, which is suitable for practical smart building operating scenarios constituting varied levels of appliance power consumption.
    \item An attention layer has been introduced into the proposed Mh-Net CNN model, which improves the model performance by reducing prediction error in the estimation of appliance power.
    \item The proposed Mh-Net CNN model has been trained with practical datasets of dynamic voltage scenarios, to improve the accuracy of the NILM technique under practical grid operating conditions.
    \item Performance of the proposed NILM scheme has been evaluated on laboratory-scale test setup under dynamic grid voltages for a large range of appliances having varied power consumptions, and also on existing UK-DALE datasets.
\end{enumerate}
\section{System Description}
\begin{figure}[h]\centering
	\includegraphics[width=0.6\linewidth]{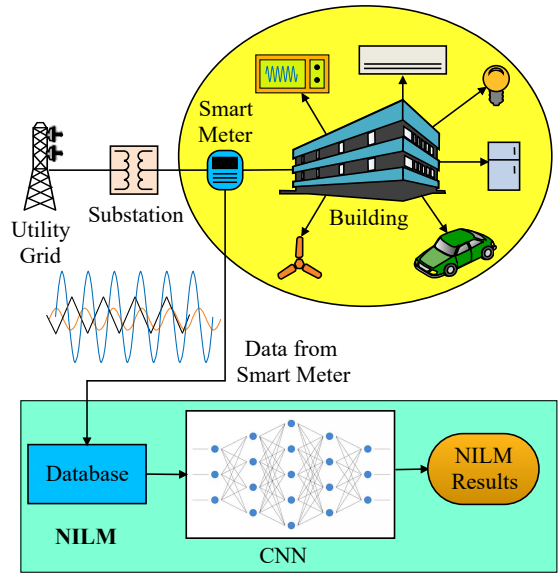}
	\caption{A schematic Diagram of a building microgrid with NILM.}
	\label{building}
	\vspace{-0.25 cm}
\end{figure}

A smart building maybe represented by a typical microgrid, primarily powered by the utility grid and may consist of on-site renewable energy sources (RES). The schematic diagram of a grid connected building microgrid is illustrated in Fig. \ref{building}. As can be observed from the figure, the building microgrid consists of a wide range of user appliances such as incandescent bulb (IB), LED tube (LT), ceiling fan (CF), air-conditioner (AC), which are connected to the utility grid through a smart meter. The smart meter records the aggregate energy consumption of the building which helps the building energy management system to monitor its consumption. For the purpose of load monitoring, an NILM technique has been proposed to perform online estimation of on/off operation and power consumption of individual appliances. Accordingly, the aggregated  data acquired by the smart meter is uploaded to a cloud-database which is further processed by the proposed Mh-Net CNN model. Individual models of each appliances are trained using the large historical aggregated and individual active \& reactive power datasets, considering dynamic voltage. The proposed Mh-Net based CNN NILM technique consists of an attention layer which improves the NILM performance and its accuracy.
\vspace{-0.3 cm}
\section{Methodology}
\vspace{-0.15 cm}
An NILM problem is first introduced in this section. Further, the propsed Mh-Net CNN model has been discussed, followed by a detailed explanation of the  attention layer introduced in the proposed model.
\vspace{-0.2 cm}
\subsection{NILM problem}
An energy meter connected between a distribution network and a building, measures the aggregated power $(y_1 ,y_2, \ldots ,y_T)$ consumed by its $N_t$ appliances. The NILM algorithms disaggregates the aggregated power into appliance level power consumption $(x^1_t, x^2_t, \ldots ,x^N_t)$ at each interval of time ($t=1,2,\ldots ,T_n$). The relation between appliance power and aggregate power is expressed as follows,
\begin{equation}
y_t = \sum_{n\in N_t}x^n_t + \varepsilon_t
\end{equation}
where, $y_t$ is the aggregate power consumed at time $t$, $x^n_t$ is the power consumed by $n^{th}$ appliance at time $t$, $N_t$ and $\varepsilon_t$ are the number of active appliances and noise term at time $t$, respectively. The function of NILM is to receive signal $y_t$ and compute signal $x^n_t$ at each time step $t$ \cite{Kelly2015NeuralND}.

\begin{figure}
    \centering
    \includegraphics[width=0.9\linewidth]{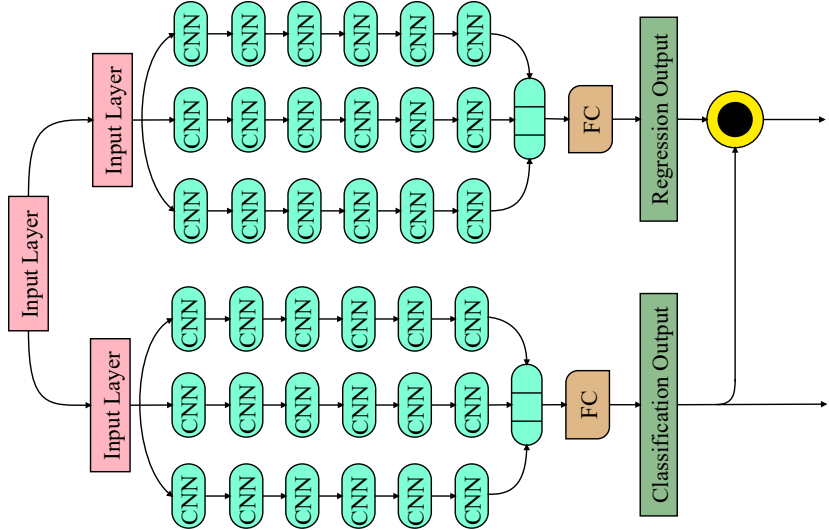}
    \caption{Proposed Mh-Net model}
    \label{Mhnetmodel}
    \vspace{-0.4cm}
\end{figure}
\subsection{Proposed Mh-Net model}
In this design, on/off operating state of appliance is directly connected into the main model through a different network \cite{en12112203}. In this work, the structural design of sub-task gated network (SGNet) has been adopted as base model.
The proposed NILM calculates the power consumption of an appliance for a sequence $l$. The input signal $(y_t,y_{t+1},\ldots, y_{t+m})$ of length $m$ is used by the NILM to calculate output signal $(x^n_{(t+m-2l)/2},x^n_{(t+m-2l+2)/2},\ldots,x^n_{(t+m+2l)/2})$ for $n^{th}$ appliance. 
The formulation of sequence to sequence machine learning (ML) model is done in two stages of operation. In first stage, the connection of number of cascaded CNN layers is done, and in second stage fully connected (FC) layer is connected to compute the task $\widehat{x}^n=f(y)$ where, $\widehat{x}^n$ is the predicted sequence \cite{article1}. To identify the on/off status of an appliance and their power consumption, two different sub network such as $f_{on/off}:R^+_{l}\rightarrow [0,1]$ and $f_{power}:R^+_{l}\rightarrow R^+$ are used for this task \cite{en12112203}. An additional auxiliary variable $(o^n_{(t+m-2l)/2},o^n_{(t+m-2l+2)/2},\ldots,o^n_{(t+m+2l)/2})$ is used to find the status (on/off) for $n^{th}$ appliance. The on/off status of appliance is presented as a probability between $\{0,1\}$, which is defined using a function $\widehat{o}^n=f(y)$. The final output of the network is defined as follows,
\begin{equation}
\widehat{x}^n=f_{on/off}^n \odot f_{power}^n
\end{equation}
The architecture of the proposed Mh-Net CNN model is shown in Fig. \ref{Mhnetmodel}. In the porposed model, initially the calculation of the different features considering the different scaling ratio in Mh-Net CNN is performed. The calculated features are passed through the attention layer to extract the important features. At last, the information is provided to the FC layer to convert this feature values into desired information. The detailed description of model is explained in succeeding subsections.

\subsection{Dilated Convolutions}
\begin{figure}[h]\centering
	\includegraphics[width=\linewidth]{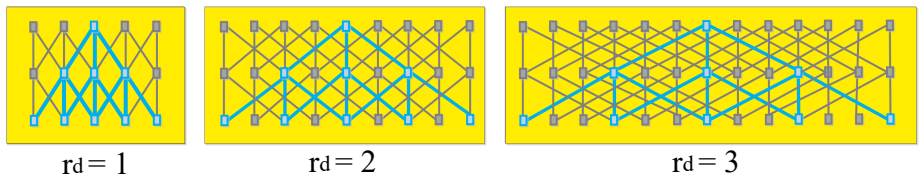}
	\caption{Explanation of input-output relation with different dilation ratio \cite{8897031}.} 
	\label{dilated}
\end{figure}

In this work, a method to incorporate the multi-scale features has been adopted. The proposed Mh-Net model has three parallel paths, considering different dilation ratios ($r_d=\{1,2,3\}$) to extract the desired information at different timescale. Finally, the information is concatenated to pass onto the next stage. The dilation ratio impacts the input and output relation, as large input with large dilation impacts the output at large dilation. Fig. \ref{dilated} illustrates the impact on output when input is passed through the three parallel CNN models with different dilation ratio. The operation is followed by predicting the appliance power consumption as well as its operating status.

\subsection{Attention Layer}
The proposed model at the concatenation stage acquires large information at different scales. The concatenated information has large information about the input signal, which is passed through the attention layer to understand the input-output relationship.\
The experimental evaluation shows that the attention based model performs better for NILM problem. The application of attention layer improves the error in predicted power and predicted on/off status of the appliance. This helps NILM process is assigning weights to each position on the concatenated signal according to their importance. Which helps the signal in activation on the position or in another word, it finds the change in operating status of appliance. This attention neural network (NN) highlights the power consumption of the appliance, which improves the performance of NILM. Therefore, the predictor can predict better than the model with absence of attention layer. Hence to improve the performance of the model, the proposed model utilizes the specific attention model whose learning function depends on the hidden state vector ($h_t$), which is calculated using NN model such as $e_t=f(h_t)$. The values obtained using the learning function is normalised using the formula given as;
\begin{equation}
    \alpha_t=\frac{exp(e_t)}{\sum_{t \in T}exp(e_t)}
\end{equation}
The output value of the attention model is calculated using the equation, $c=\sum_{t\in T}\alpha_t*h_t$. The calculated values are passed through the fully connected layer to obtain the desired output.
\subsection{Performance Metrics}
To assess the accuracy of the proposed model several performance metrics have been considered namely, mean of absolute error (MAE), signal aggregate error (SAE) and F1 score \cite{8897031}. Mathematical representation to evaluate MAE and SAE is given as follows;
\begin{equation}
    MAE= \frac{\sum_{i=1}^{n} |P_{a,i}-P_{p,i}|}{n}
\vspace{-0.1cm}
\end{equation}
\begin{equation}
    SAE= \frac{1}{n} {\sum_{i=1}^{n}} \frac{1}{m} |P_{a,i}-\hat P _{p,i}|
\vspace{-0.1cm}
\end{equation}
where, {$P_a$} is actual power consumed by an appliance, {$P_p$} is the predicted power consumption of the appliance, \textit{n} is the number of disjoint time periods with length \textit{m} and {$\hat P_{p,i}$} is aggregated predicted power of appliances.\\ 
\indent Another performance metric is the F1 score, which measures the model's accuracy and is calculated using harmonic mean of the precision ($P_r$) and recall (R). $P_r$ refers to the fraction value of the positive values among the values which are classified as positive by the model. Further, R is the sensitivity and is calculated as a fraction of positive classified values, among the total positive values. 

Mathematically, \textit{F1} score is calculated as follows,
\begin{equation}
    F1=\frac{2.{P_r}.R}{{P_r}+R}
\end{equation}
The values of \textit{P} and \textit{R} are calculated as follows, 
\begin{equation}
    P_r=\frac{TPV}{TPV+FPV}
\end{equation}
\begin{equation}
    R=\frac{TPV}{TPV+FNV}
\end{equation}
where, \textit{TPV}, \textit{FPV} and \textit{FNV} are true positive value, false positive value and false negative value, respectively. \\

\section{Results and Discussion}
In this section, performance of the proposed Mh-Net CNN model is evaluated on the laboratory test setup under different operating conditions, and also on the widely used UK-DALE dataset. A wide range of appliances are considered and the performance of NILM is investigated using performance metrics such as MAE, SAE, F-1 score. 
\subsection{Test Setup.}
\begin{figure}
    \centering
    \includegraphics[width=0.95\linewidth]{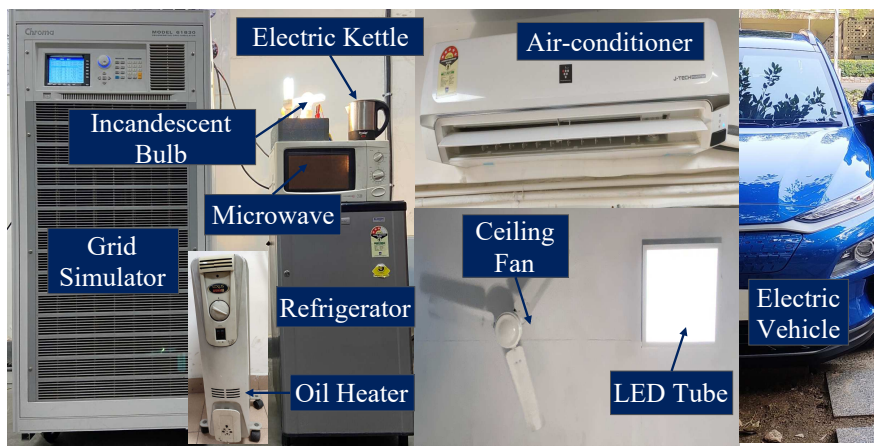}
    \caption{The experimental laboratory test setup.}
    \label{Setup}
\end{figure}

An experimental laboratory test setup has been developed to analyze appliance behaviour at dynamic voltage scenarios as shown in Fig. \ref{Setup}. The setup consists of a Chroma-make Regenerative Grid Simulator–61800 for generating different voltage scenarios. The setup compromises of nine single phase appliances, the details of which are listed in Table \ref{appliances}. 
\begin{table}[h]
\caption{Details of appliances used in laboratory setup.}\label{appliances}\centering
    \begin{tabular}{c|c|c}
        \hline
        \hline
        Name of Appliance & Type & Specifications \\
        \hline
        \rule{0pt}{2ex} Electric Vehicle (EV) & B,D & 230V, 50Hz, 12A\\
        \hline
        \rule{0pt}{2ex} Refrigerator (RF)  & A,B & 220-240V, 50Hz, 150 litres \\
        \hline
        \rule{0pt}{2ex} Microwave (MW) & C,D & 230V, 50Hz, 1000W \\
        \hline
        \rule{0pt}{2ex} Electric Kettle (EK) & B & 230V AC, 50Hz, 1350W \\
        \hline
        \rule{0pt}{2ex} Oil Heater (OH) & B,D & 230V, 50Hz, 1500W \\
        \hline
        \rule{0pt}{2ex} Incandescent Bulb (IB) & B  & 230V, 100W  \\
        \hline
        \rule{0pt}{2ex} Air-conditioner (AC) &  C,D  & 230V, 50Hz, 2530W  \\
        \hline
        \rule{0pt}{2ex} Ceiling Fan (CF) & B,D & 220V, 50Hz, 70W   \\
        \hline
        \rule{0pt}{2ex} LED Tube (LT) & B & 220-240V, 50-60Hz, 34W  \\
         \hline
        \multicolumn{3}{c}{A= always on; B= on-off; C= continuous variable; D= multi-state} \\
        \hline
        \hline
    \end{tabular}
    \vspace{-0.3cm}
\end{table}
\\ \indent The electrical power is fed to the appliances through the grid simulator and electrical parameters data such as voltage (V), current (I), active power (P), reactive power (Q), apparent power (S), power factor (p.f.), active and reactive energy are recorded through the grid simulator at a frequency of 1 Hz. 
\subsection{Laboratory testing of appliances under dynamic operating voltage conditions}
\begin{figure*}\centering
\begin{minipage}{0.45\textwidth}
    \includegraphics[width=\linewidth]{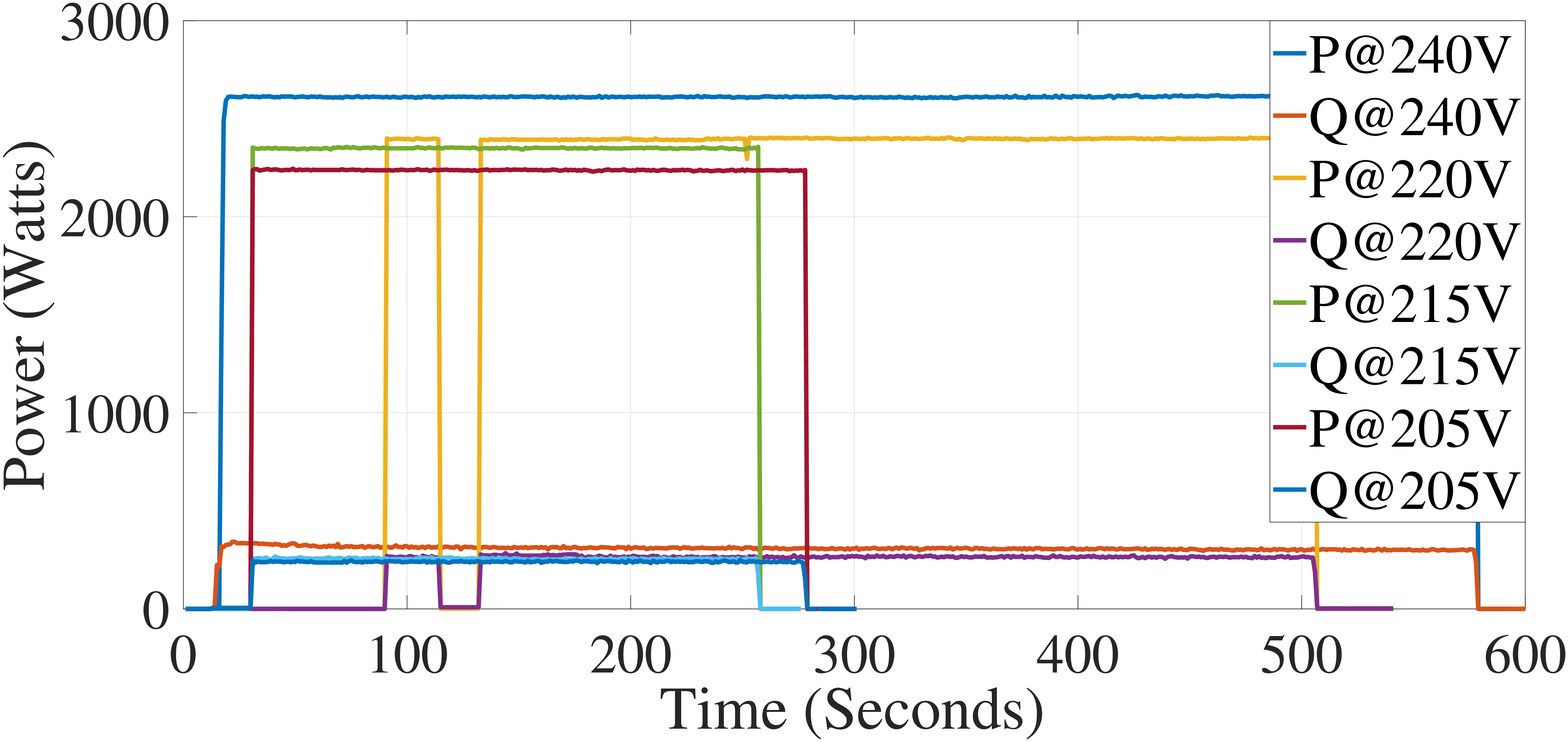}
    \subcaption{Electric vehicle}
    \label{EV3}
\end{minipage}
\begin{minipage}{0.45\textwidth}
    \includegraphics[width=\linewidth]{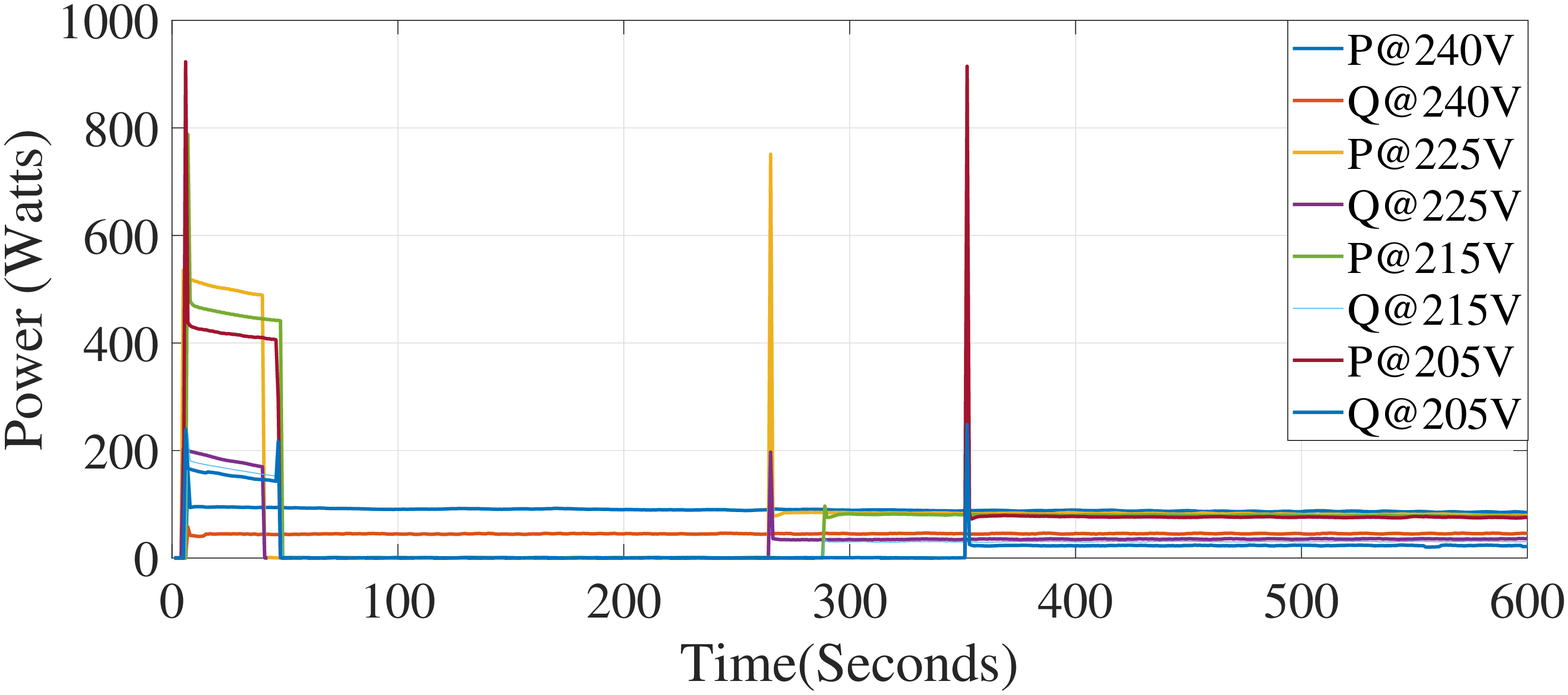}
    \subcaption{Refrigerator}    
\label{fridge}
\end{minipage}
\begin{minipage}{0.45\textwidth}
    \includegraphics[width=\linewidth]{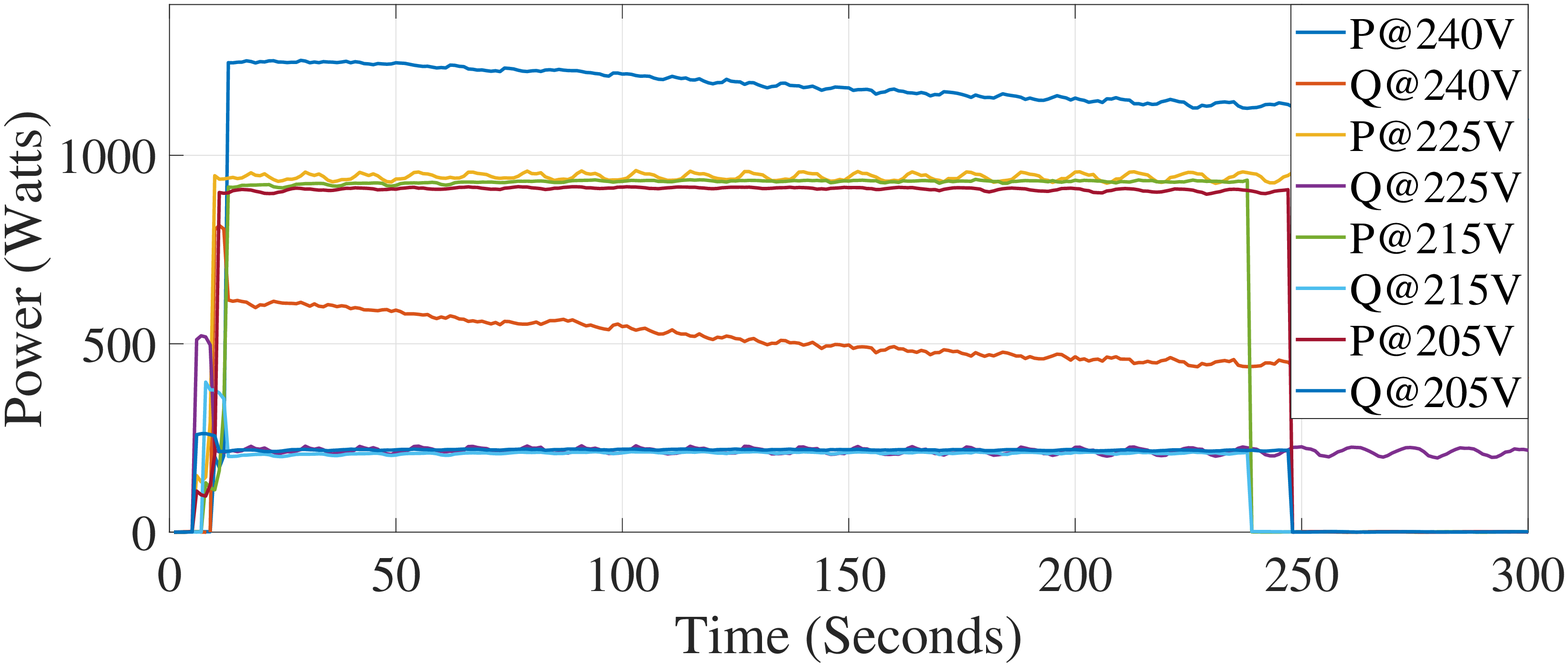}
    \subcaption{Microwave}    
\label{microwave}
\end{minipage}
\begin{minipage}{0.45\textwidth}
    \includegraphics[width=\linewidth]{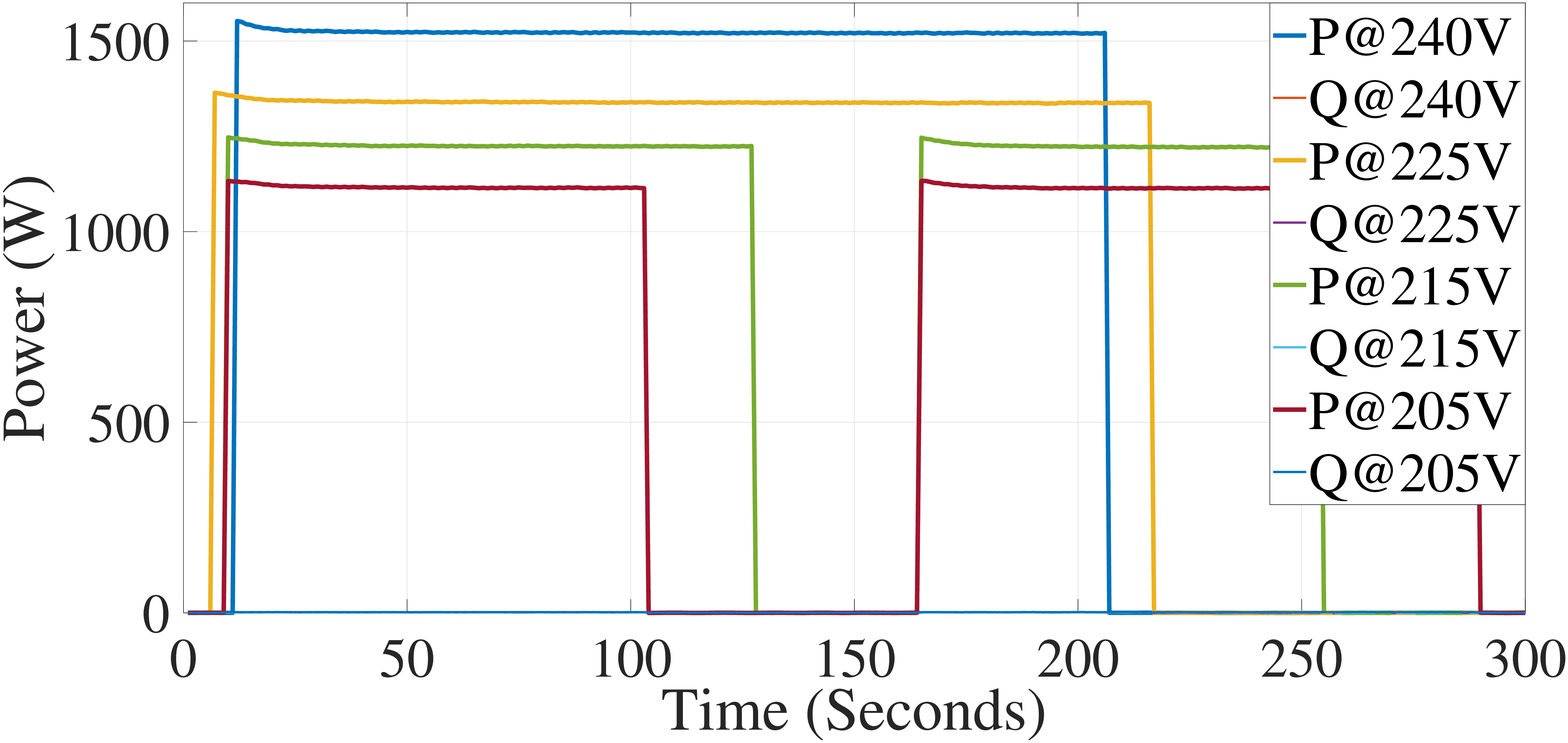}
    \subcaption{Electric kettle}
    \label{kettle}
\end{minipage}
\begin{minipage}{0.45\textwidth}
    \includegraphics[width=\linewidth]{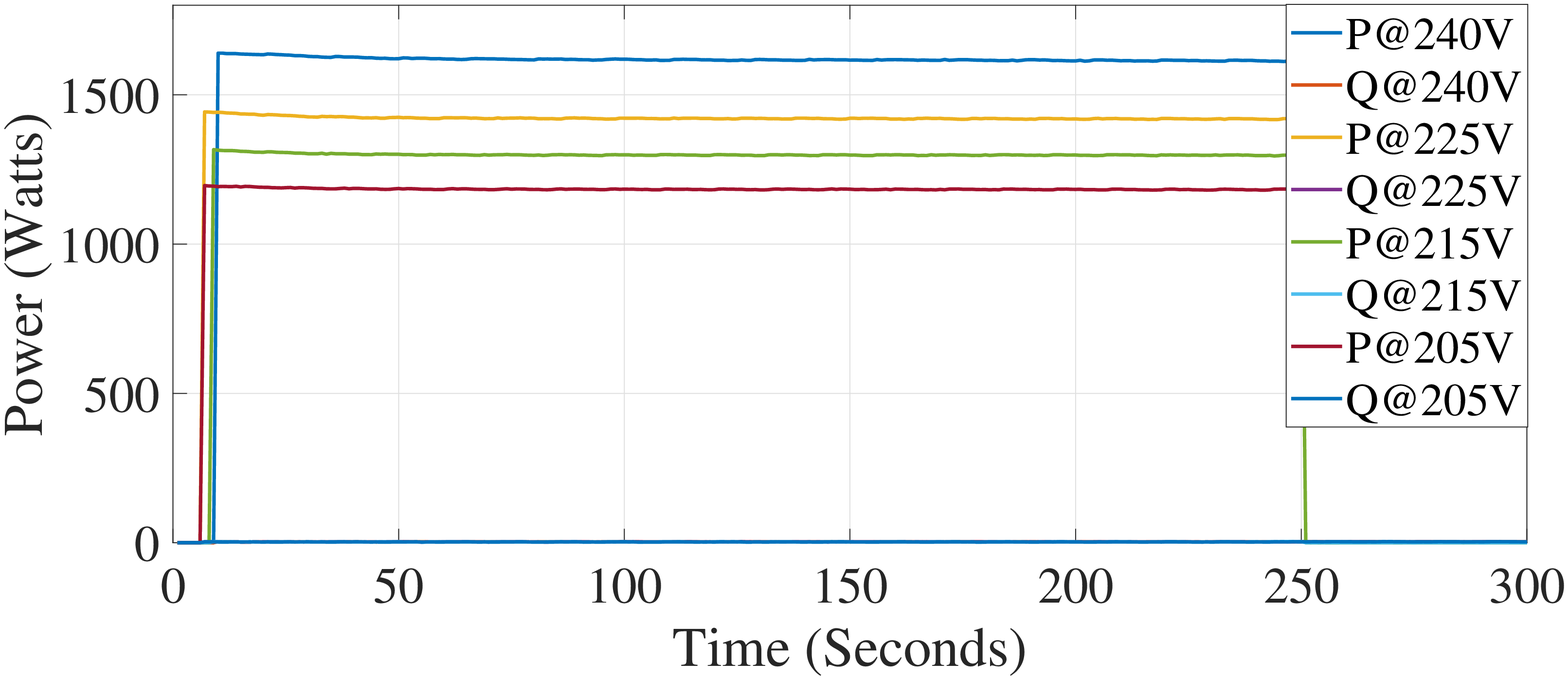}
    \subcaption{Oil heater}    
\label{oilheater}
\end{minipage}
\begin{minipage}{0.45\textwidth}
    \includegraphics[width=\linewidth]{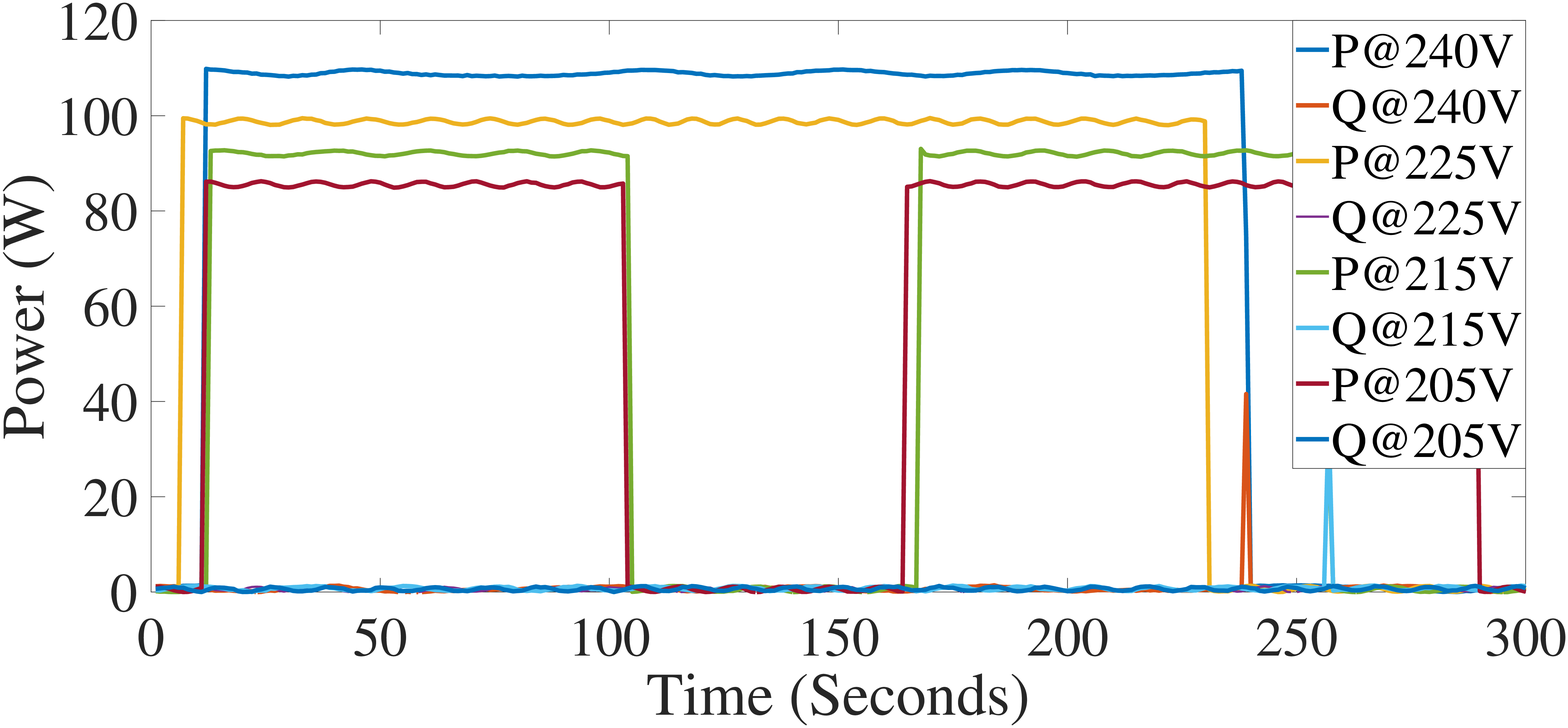}
    \subcaption{Incandescent bulb}
    \label{incandescentbulb}
\end{minipage}
\begin{minipage}{0.45\textwidth}
    \includegraphics[width=\linewidth]{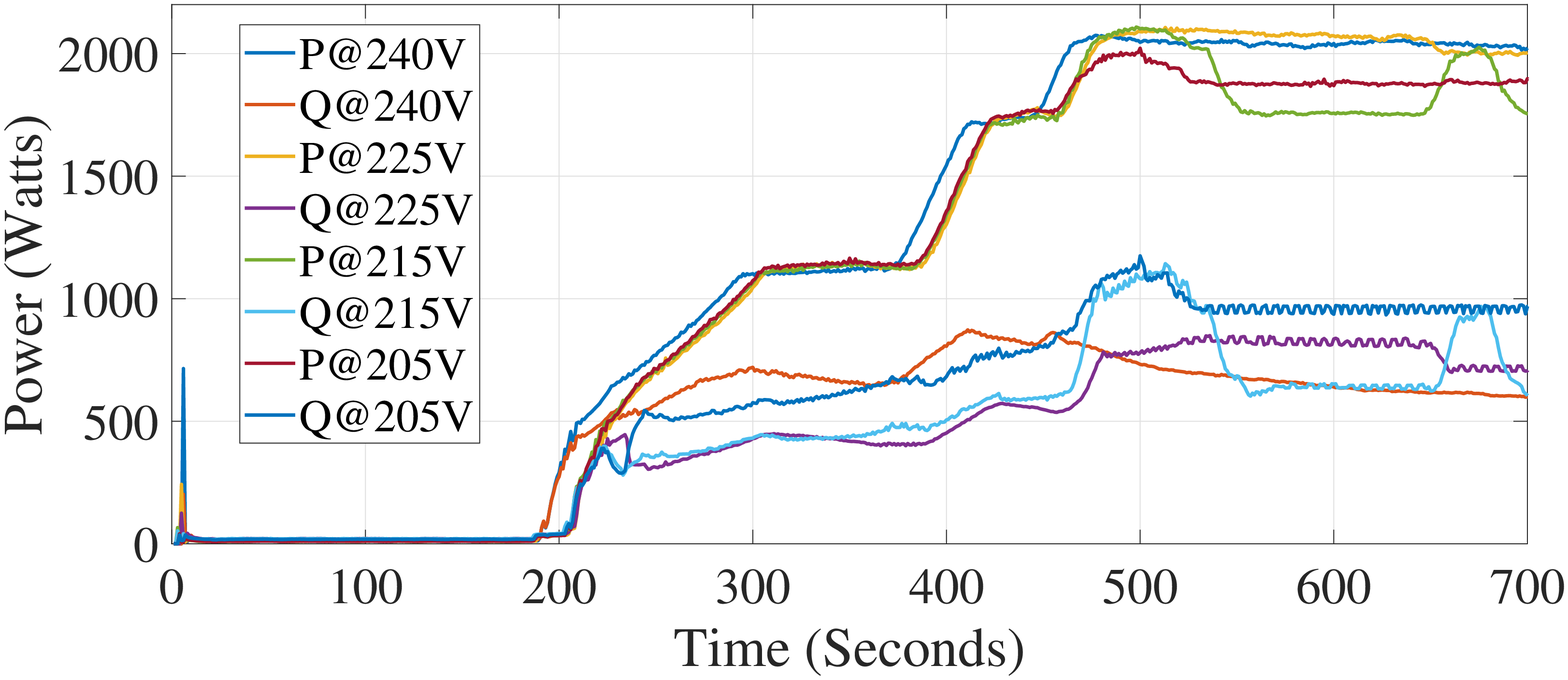}
    \subcaption{Inverter-based air conditioner}
    \label{SplitAC}
\end{minipage}
\begin{minipage}{0.45\textwidth}
    \includegraphics[width=\linewidth]{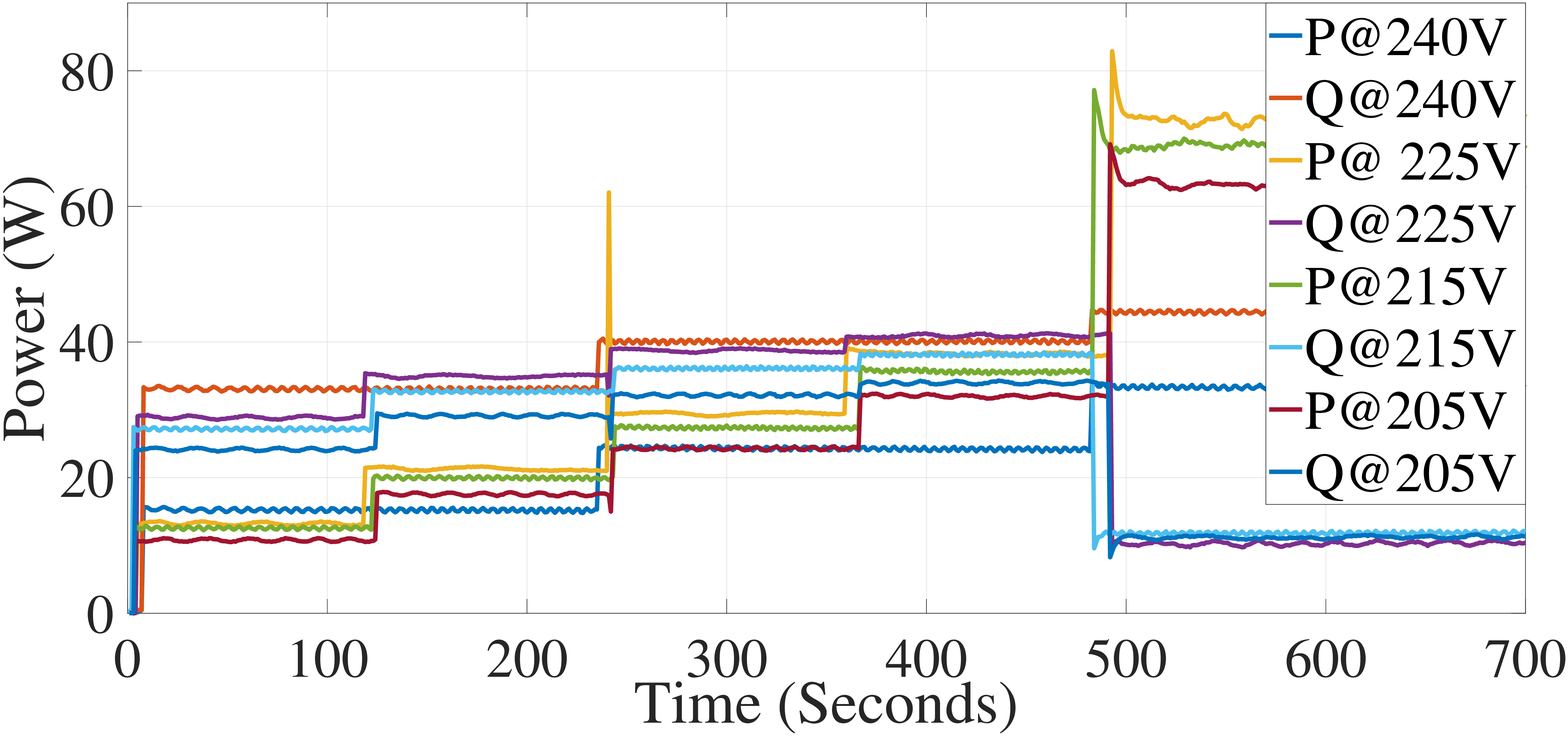}
    \subcaption{Ceiling fan}    
\label{ceilingfan}
\end{minipage}
\begin{minipage}{0.45\textwidth}
    \includegraphics[width=\linewidth]{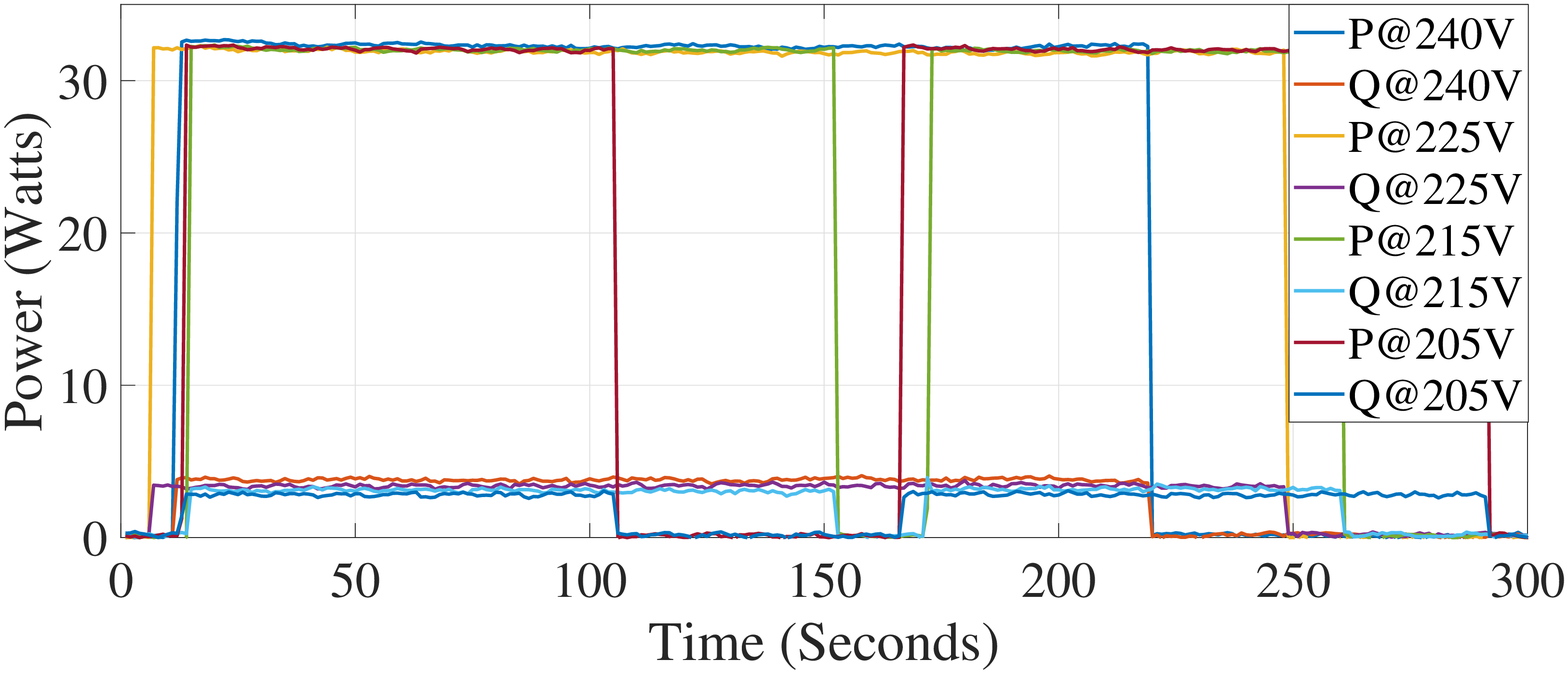}
    \subcaption{LED tube}
    \label{LEDTube}
\end{minipage}
\caption{Real and reactive power profiles from experimental testing of appliances at different voltage conditions.}
\end{figure*}

The focus of this test is to analyze the behaviour of different appliances under dynamic operating voltages, which is a practical scenario of the power distribution system. Accordingly, the grid simulator generates different voltage scenarios to emulate dynamic grid voltage conditions. The experiment has been carried out at different grid voltages which is varied within an acceptable distribution voltage range i.e. 205V to 240V @ 50Hz. The test lasted for a few minutes, considering different operating conditions of appliances such as on-off, continuous variable, multi-state, always-on. The active and reactive power profiles of EV under different voltage scenarios are shown in Fig. \ref{EV3}.
It can be observed from the figure that EV demonstrates a significant variation in active power with different voltage scenarios, but shows less variation in reactive power. Similarly, active and reactive power variations are observed for RF and MW under different voltage scenarios, as depicted in Fig. \ref{fridge} and \ref{microwave}. Voltage dependency is majorly observed in case of resistive appliance, as the current drawn by the resistive appliances varies proportionally to the change in voltage. Figs. \ref{kettle}-\ref{incandescentbulb}, show the active and reactive power profiles of EK, OH and IB, respectively. It can be observed from the figures that EK, OH and IB demonstrate a large variation in active power. No change in reactive power is observed as these appliances operate at unity power factor under different voltage scenarios. Similarly,  Fig. \ref{SplitAC} shows the active \& reactive power profiles of an inverter-based AC, and voltage dependency is observed due to variation in active and reactive powers for different voltage scenarios. Further, the active and reactive power profiles of an induction machine-based CF is depicted in Fig. \ref{ceilingfan}. The appliance CF is operated at different operating conditions such as different speeds and voltage scenarios. It has been observed that the steady-state as well as transient performance of active and reactive power of CF varies with different operating speed and voltages. 
Fig. \ref{LEDTube} depicts the power profile of a LT. However, it is observed that the active and reactive power consumption of LT are nearly uniform under different voltage scenarios. 
Hence, it has been observed from experimental evaluation that the active and reactive power profiles of major commercial appliances vary with variation in supply voltage. Therefore, for training supervised NILM models, load profiles of different commercial appliances at dynamic grid conditions must be incorporated for efficient and accurate operation of NILM. 
\subsection{Performance analysis of proposed Mh-Net CNN NILM under constant voltage scenario.}
A laboratory experiment has been performed to validate the efficacy of the proposed Mh-Net CNN model trained at constant voltage, under static grid conditions. Nine test scenarios have been simulated, each having a different combination of operating appliances. The test scenarios have been generated by operating appliances having varied levels of power consumption at different operating conditions such as speed change, multiple states, on-off etc. The individual appliance models have been trained by providing aggregated active and reactive power datsets and individual appliance's active and reactive power datasets, at a frequency of 1 Hz under constant supply voltage. For performance evaluation of the proposed NILM model, aggregated active and reactive power datasets under the different test scenarios are provided to the individual models. It is observed that the NILM model trained at constant voltage condition, successfully extracts the individual load active power and ToU of each appliance. Aggregated active and reactive power profiles under the different test scenarios, with comparison of actual and measured appliance active power are shown in Fig. \ref{CL1}-\ref{CL9}. Moreover, the identified ToU of the appliances is shown using predictability. Performance metrics of the proposed model for different appliances under constant voltage is shown in Table \ref{step}. Results reveal that the proposed model under constant voltage provides higher accuracy even for random combinations of appliances having wide range of power consumption levels. 
\begin{figure*}\centering
\vspace{-1.5cm}
\begin{minipage}{0.45\textwidth}
\begin{subfigure}[b]{\textwidth}
     \includegraphics[width=\linewidth]{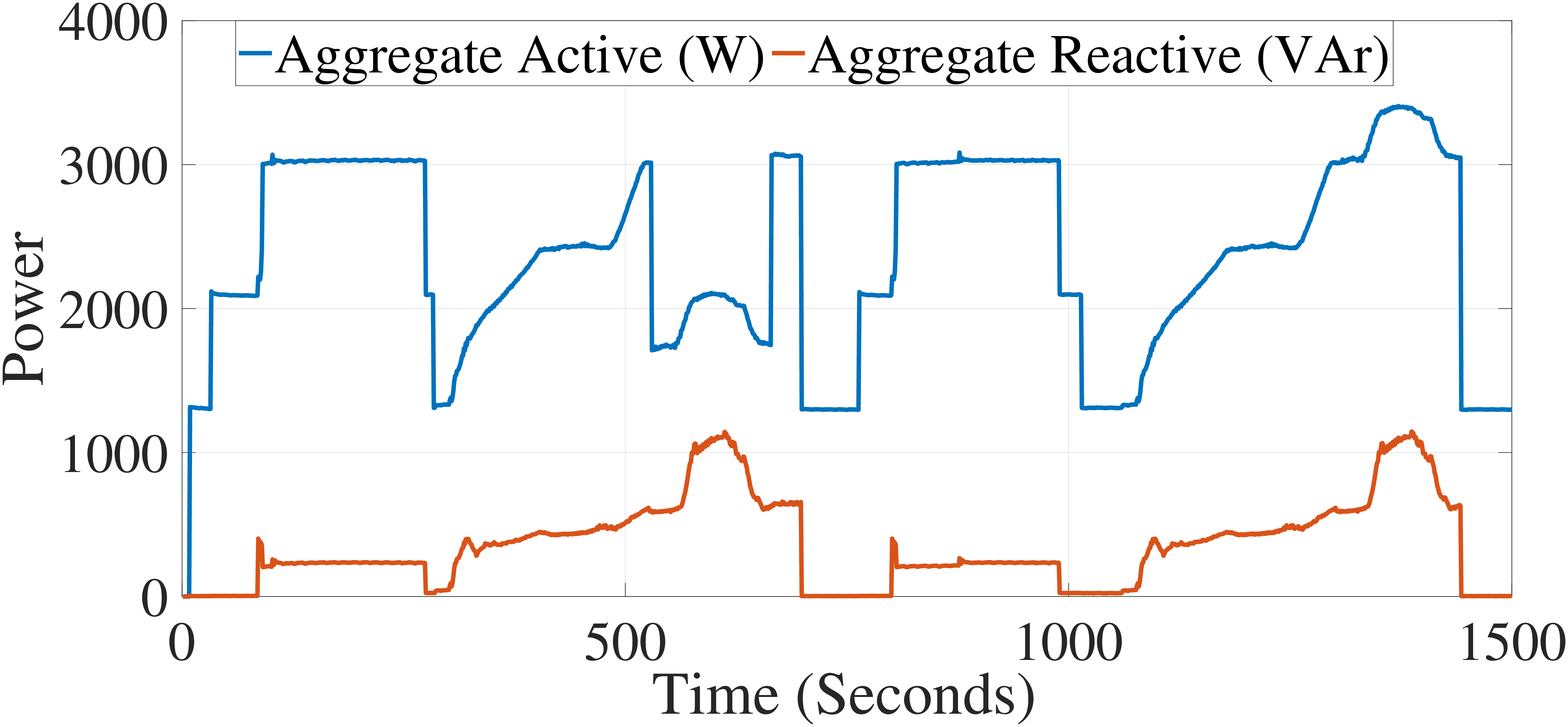}
\end{subfigure}
\begin{subfigure}[b]{\textwidth}
\includegraphics[width=\linewidth]{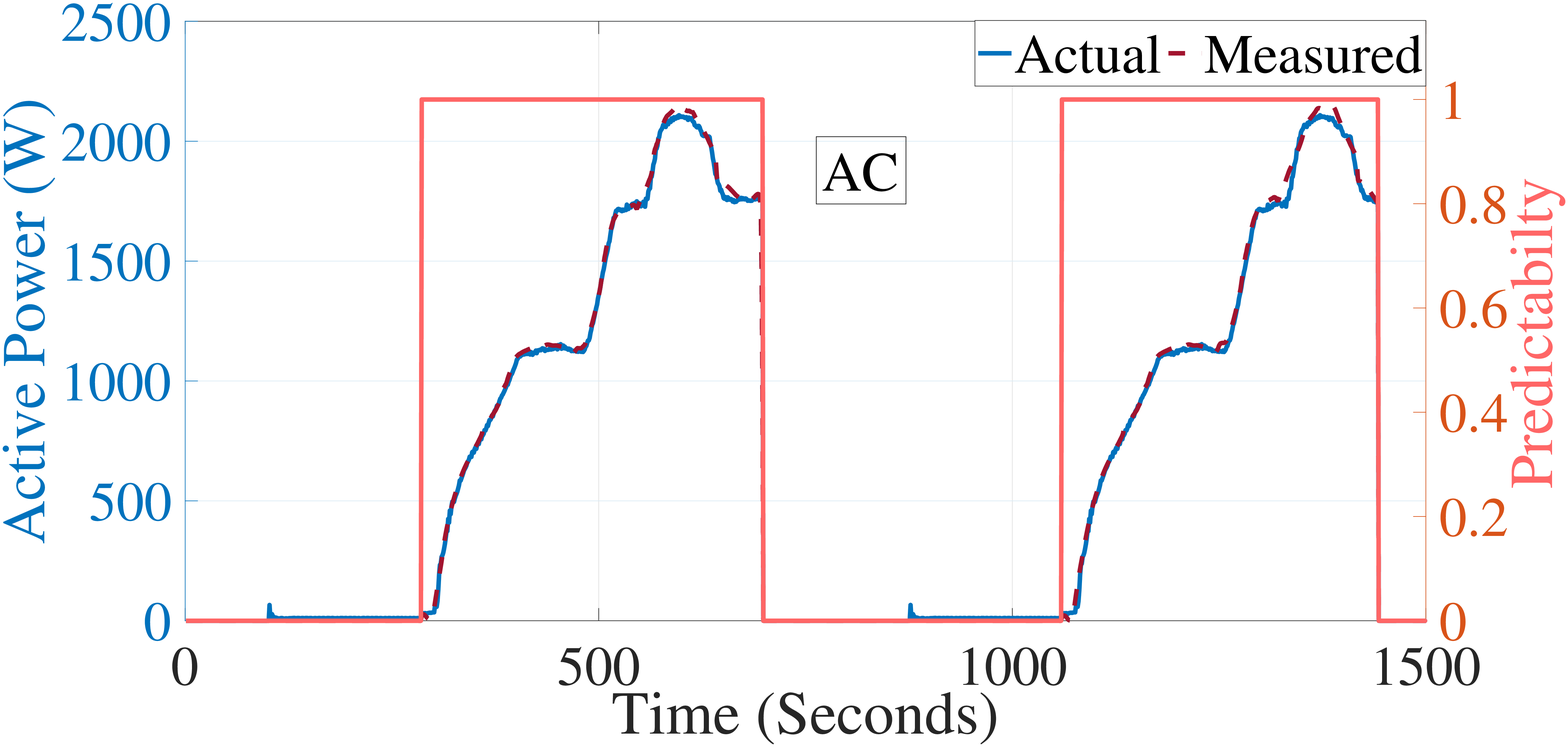}
\subcaption{Load profile of an air-conditioner}  \label{CL1}
\end{subfigure}
\end{minipage}
\begin{minipage}{0.45\textwidth}
\begin{subfigure}[b]{\textwidth}
\includegraphics[width=\linewidth]{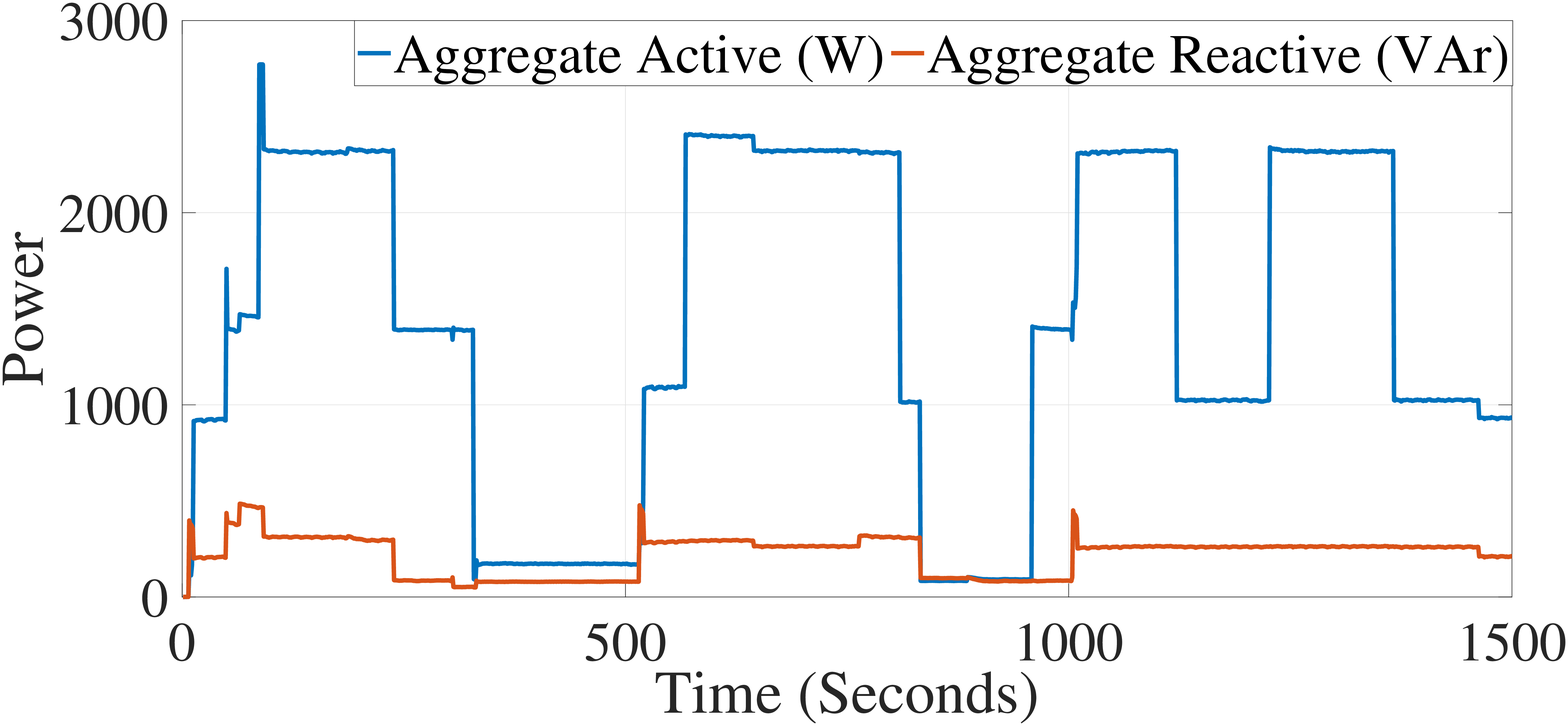}
\end{subfigure}
\begin{subfigure}[b]{\textwidth}
 \includegraphics[width=\linewidth]{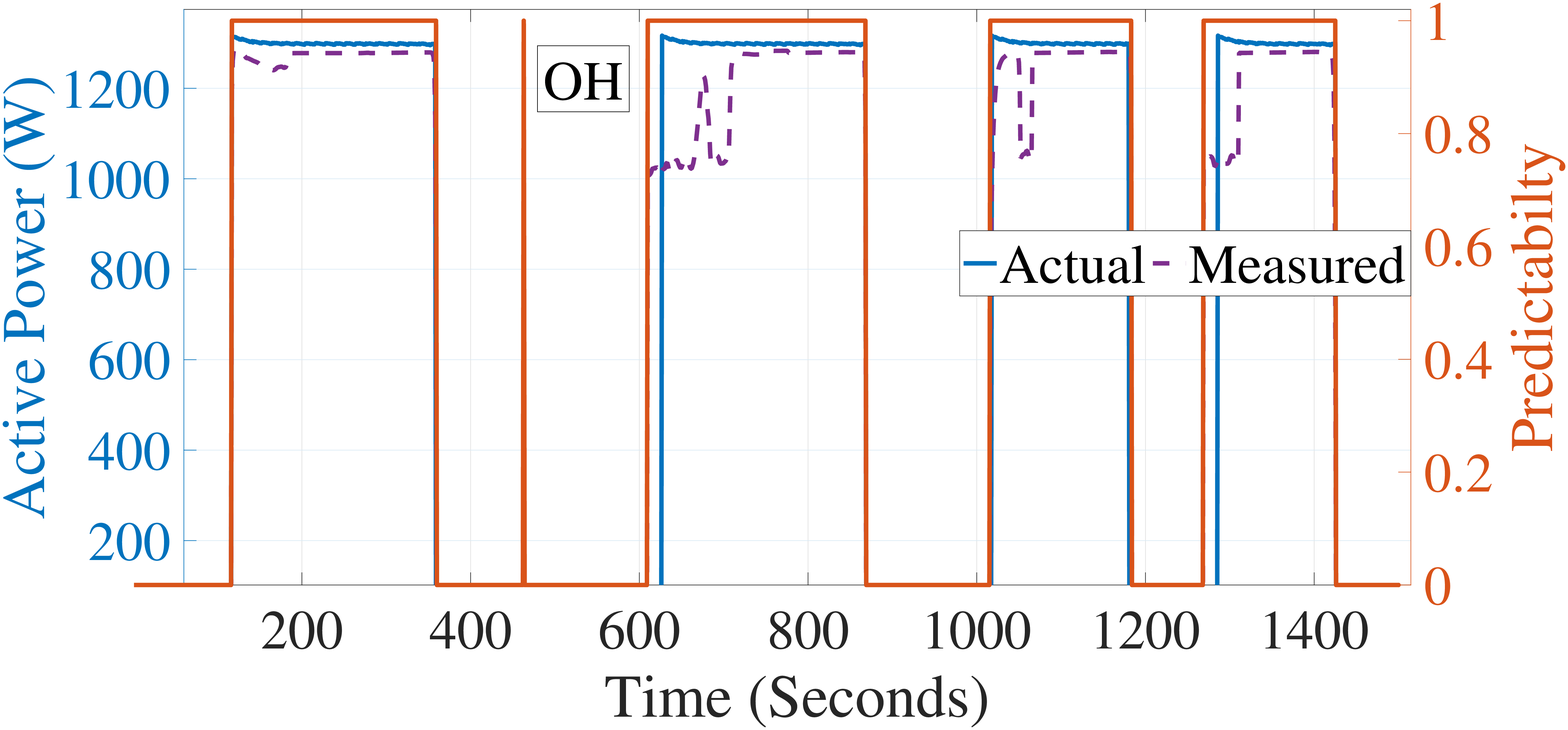}
\subcaption{Load profile of an oil heater}  \label{CL2}
\end{subfigure}
\end{minipage}
\begin{minipage}{0.45\textwidth}
\begin{subfigure}[b]{\textwidth}
     \includegraphics[width=\linewidth]{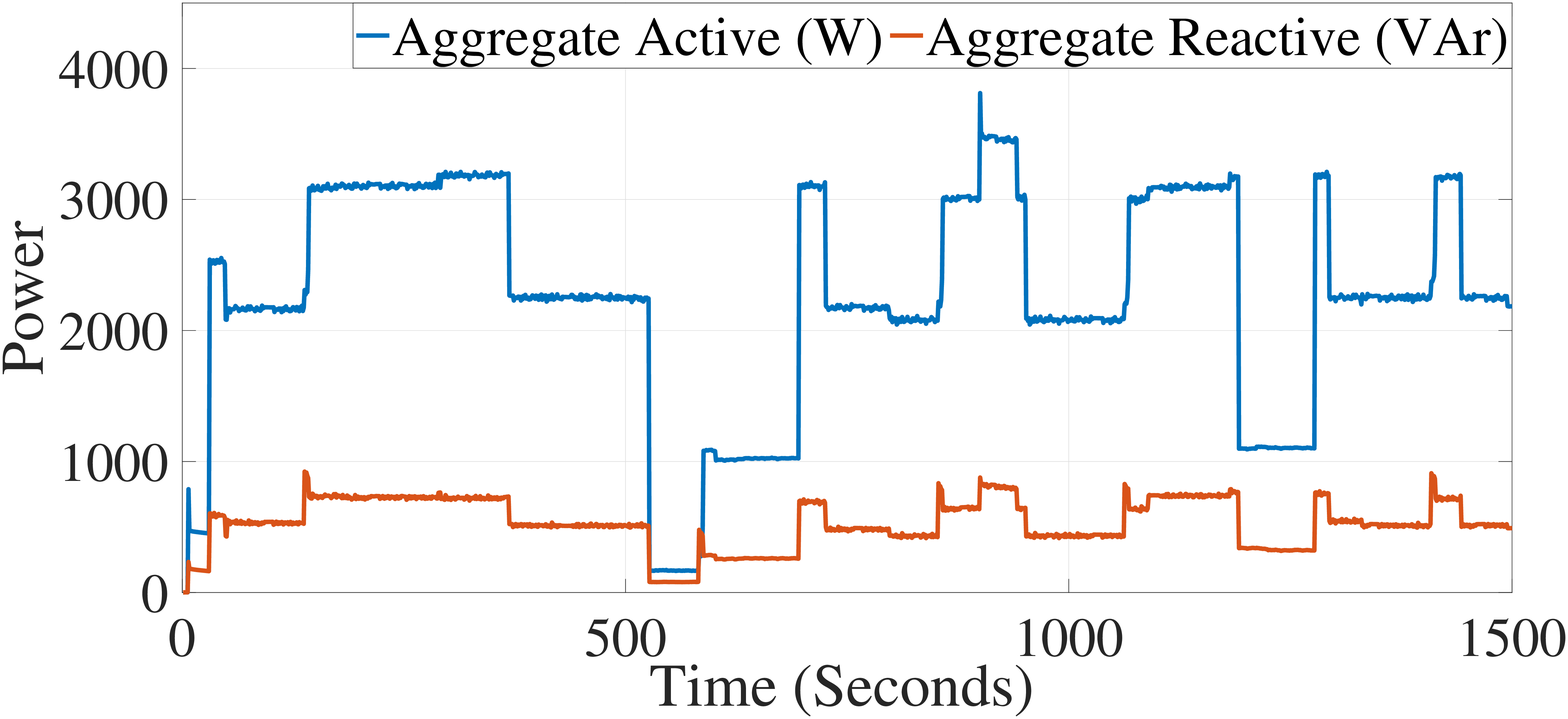}
\end{subfigure}
\begin{subfigure}[b]{\textwidth}
     \includegraphics[width=\linewidth]{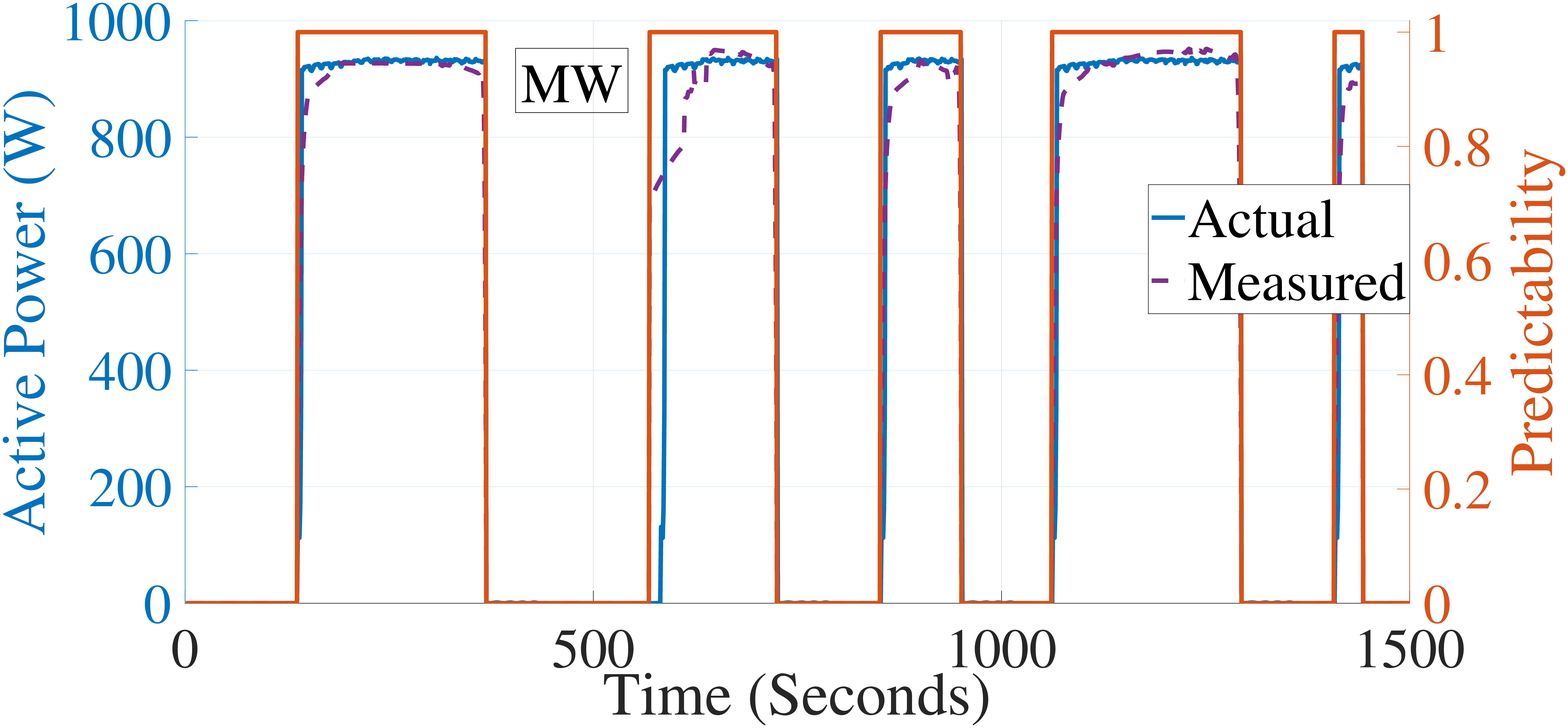}
  \subcaption{Load profile of a microwave}  \label{CL3}
\end{subfigure}
 \end{minipage}
\begin{minipage}{0.45\textwidth}
\begin{subfigure}[b]{\textwidth}
\includegraphics[width=\linewidth]{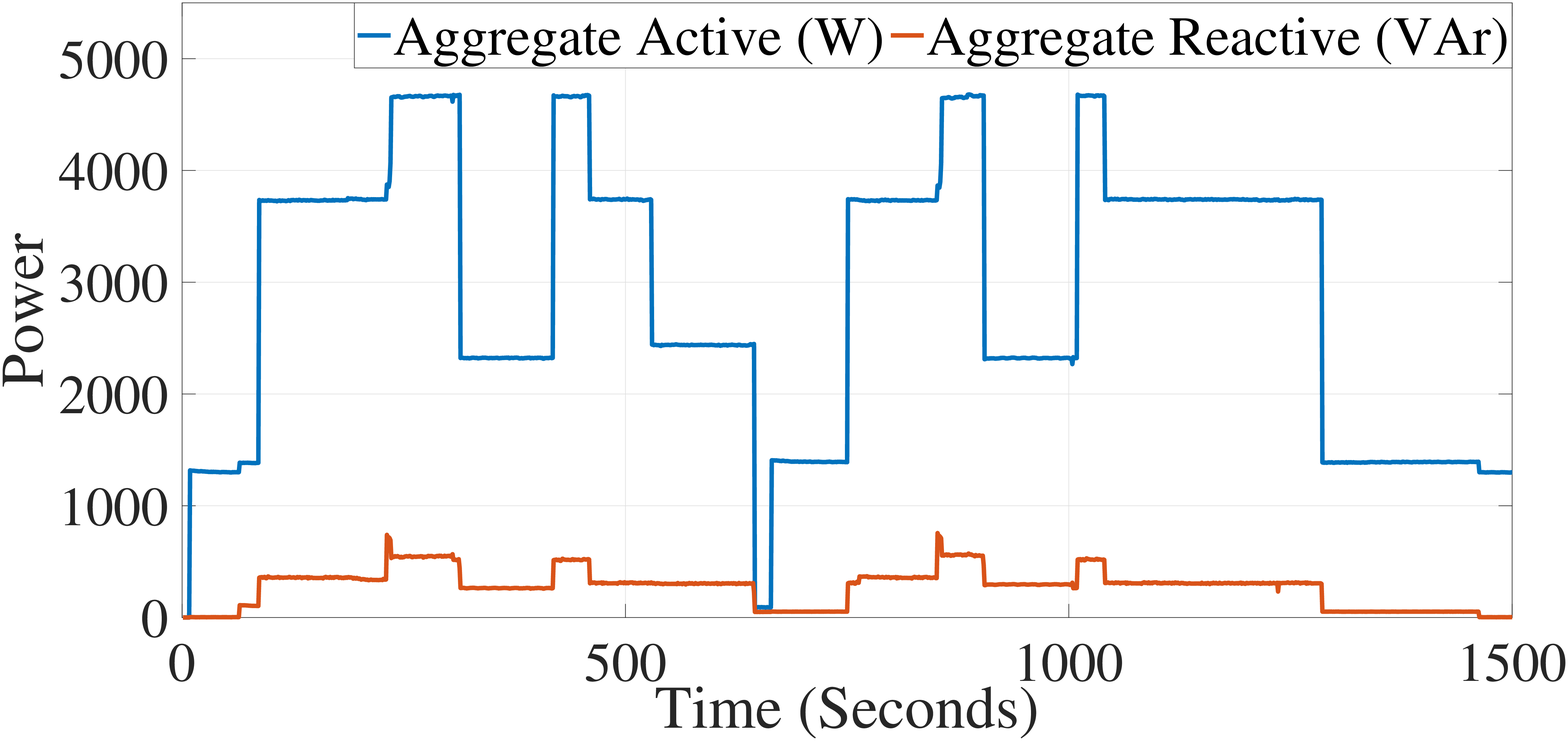}
\end{subfigure}
\begin{subfigure}[b]{\textwidth}
\includegraphics[width=\linewidth]{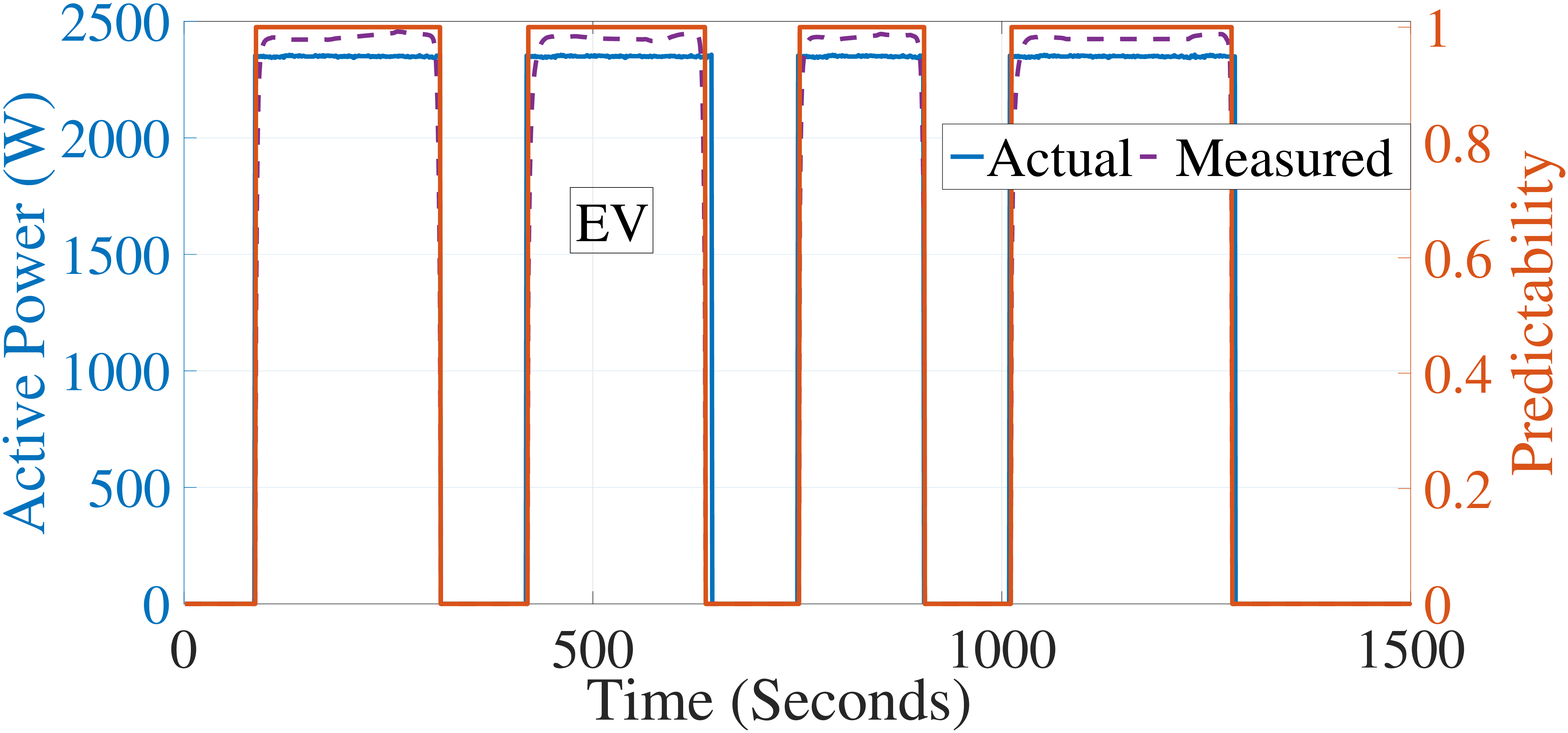}
\subcaption{Load profile of an electric vehicle}  \label{CL4}
\end{subfigure}
\end{minipage}
\begin{minipage}{0.45\textwidth}
\begin{subfigure}[b]{\textwidth}
\includegraphics[width=\linewidth]{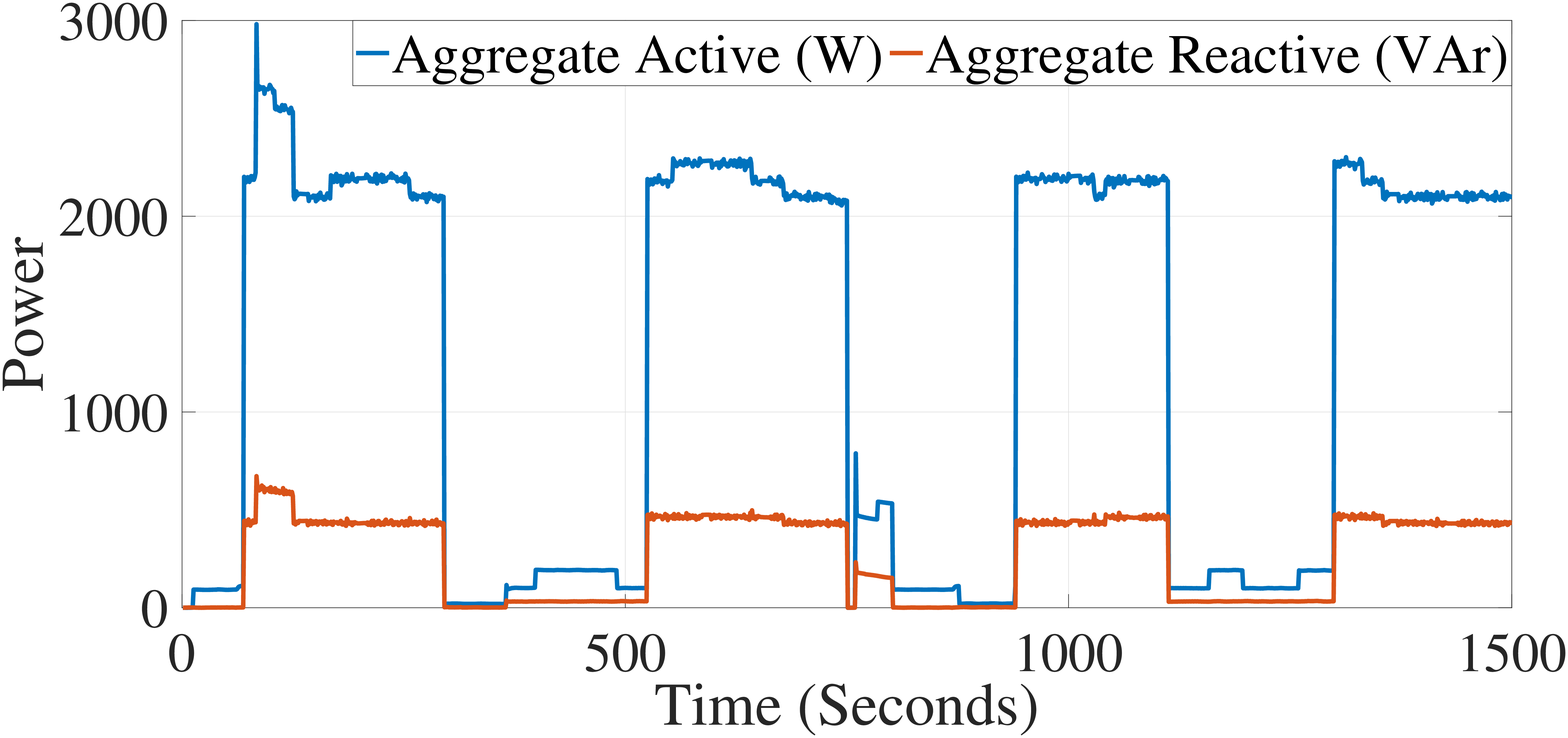}
\end{subfigure}
\begin{subfigure}[b]{\textwidth}
\includegraphics[width=\linewidth]{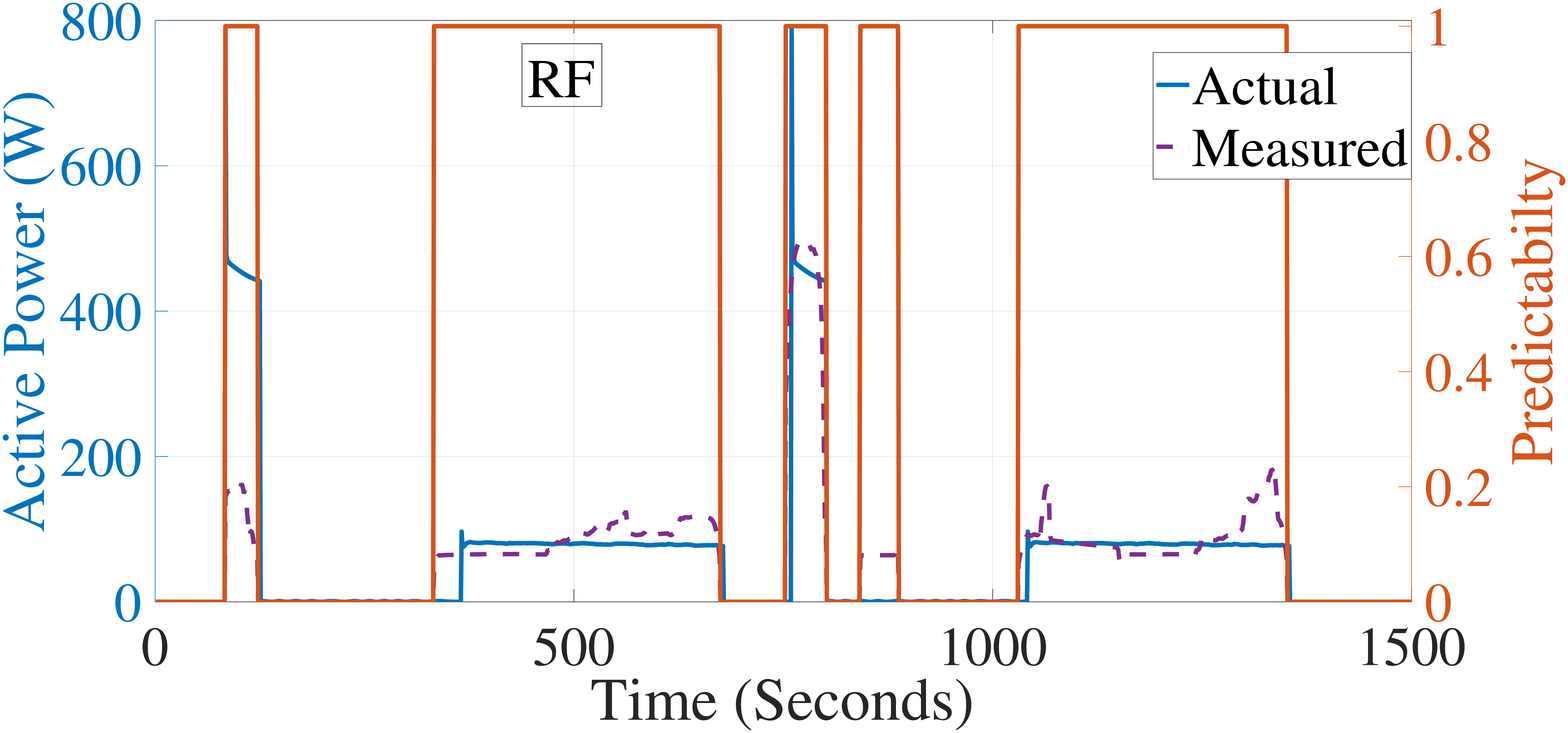}
\subcaption{Load profile of a refrigerator}  \label{CL5}
\end{subfigure}
\end{minipage}
\begin{minipage}{0.45\textwidth}
\begin{subfigure}[b]{\textwidth}
\includegraphics[width=\linewidth]{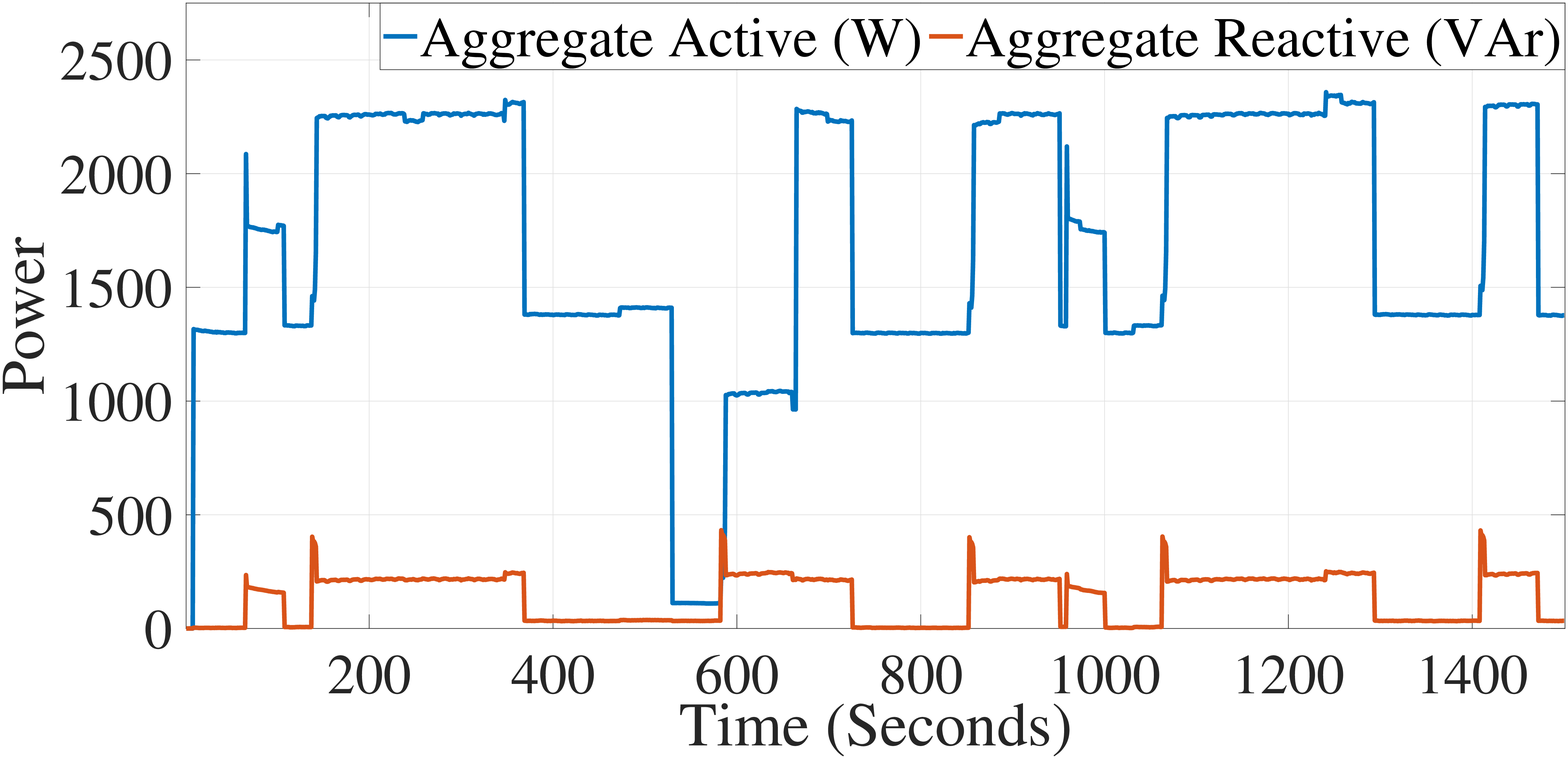}
\end{subfigure}
\begin{subfigure}[b]{\textwidth}
\includegraphics[width=\linewidth]{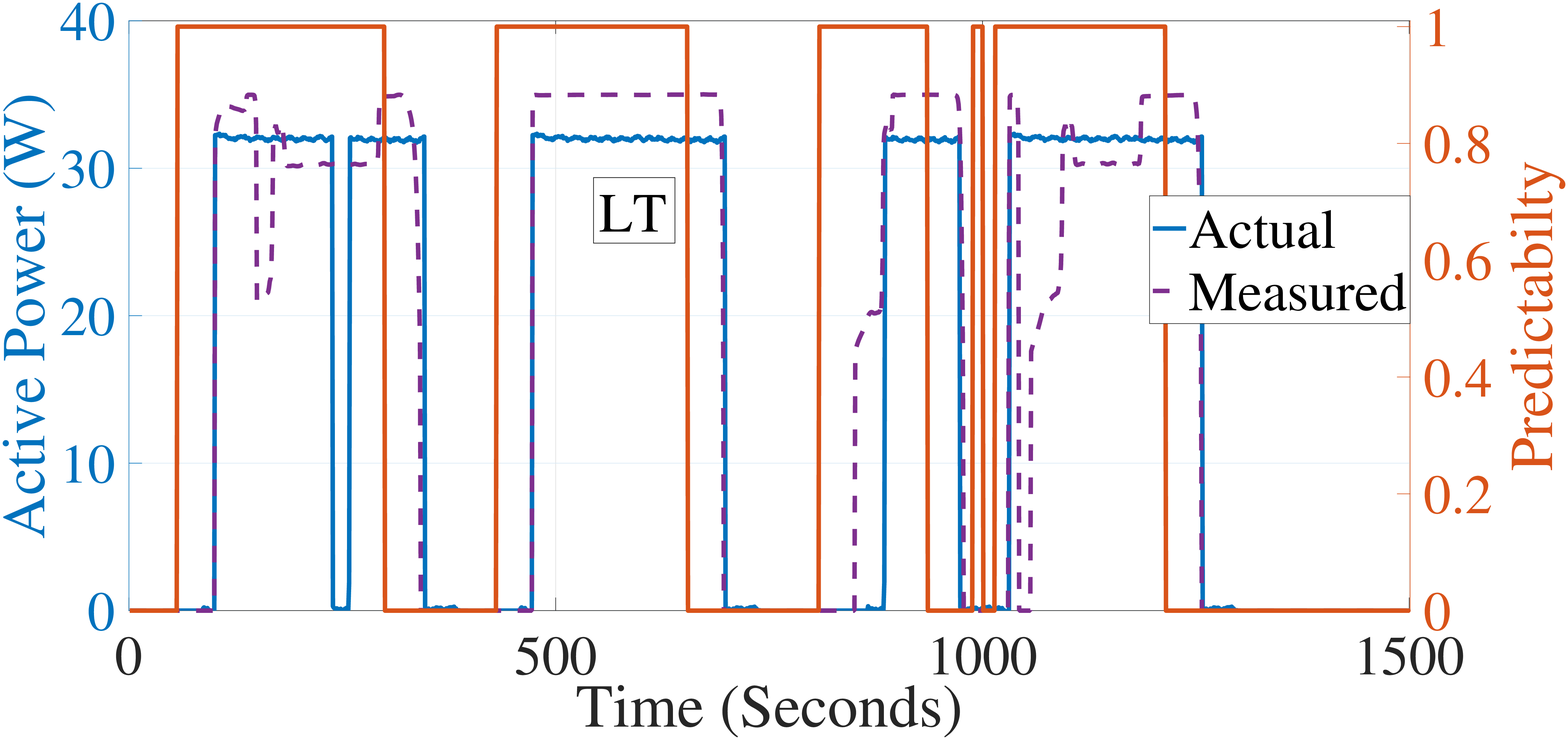}
\subcaption{Load profile of a LED tube}  \label{CL6}
\end{subfigure}
\end{minipage}
\caption{Performance of proposed Mh-Net NILM model under constant supply voltage.}
\end{figure*}
\begin{figure*}
\ContinuedFloat
\begin{minipage}{0.45\textwidth}
\begin{subfigure}[b]{\textwidth}
\includegraphics[width=\linewidth]{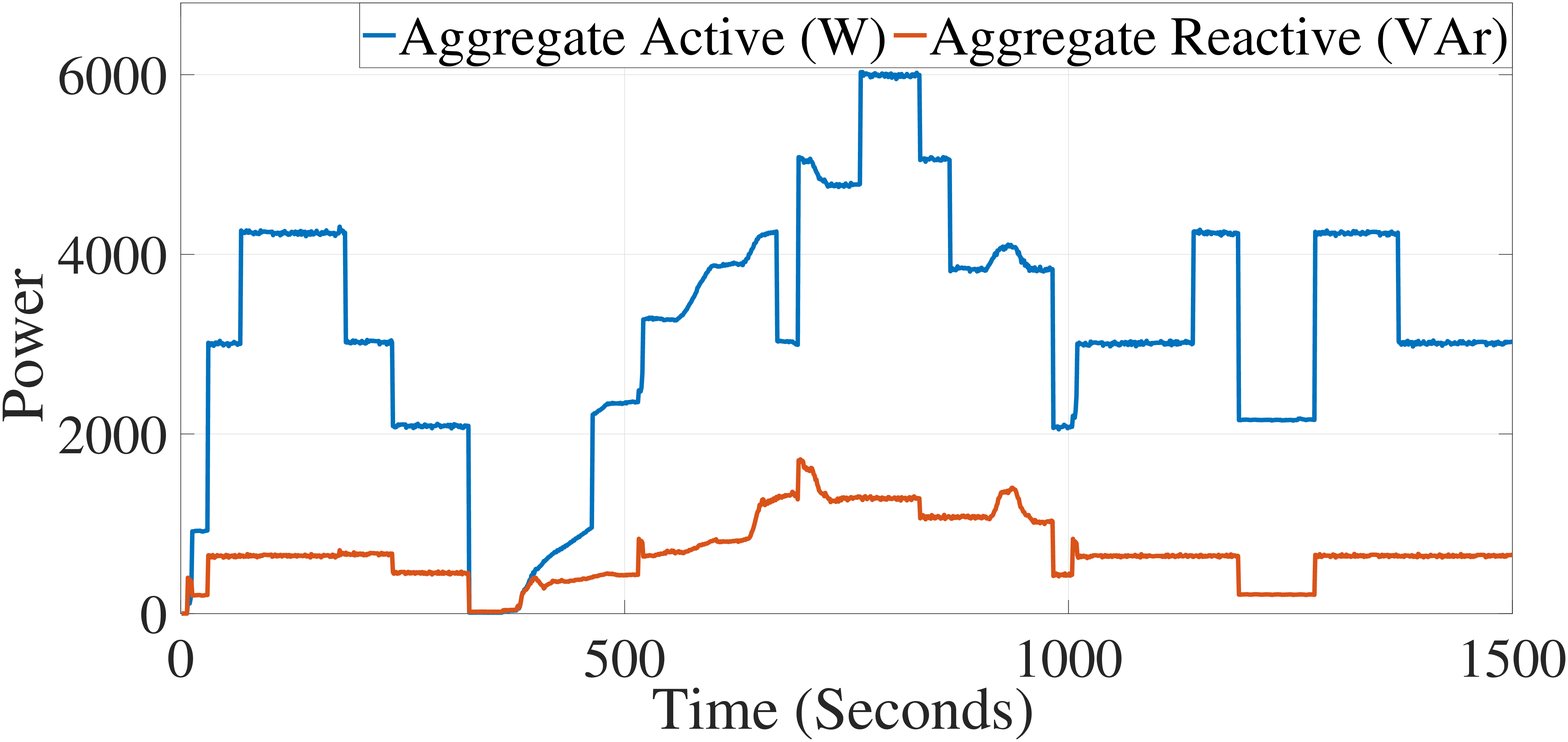}
\end{subfigure}
\begin{subfigure}[b]{\textwidth}
\includegraphics[width=\linewidth]{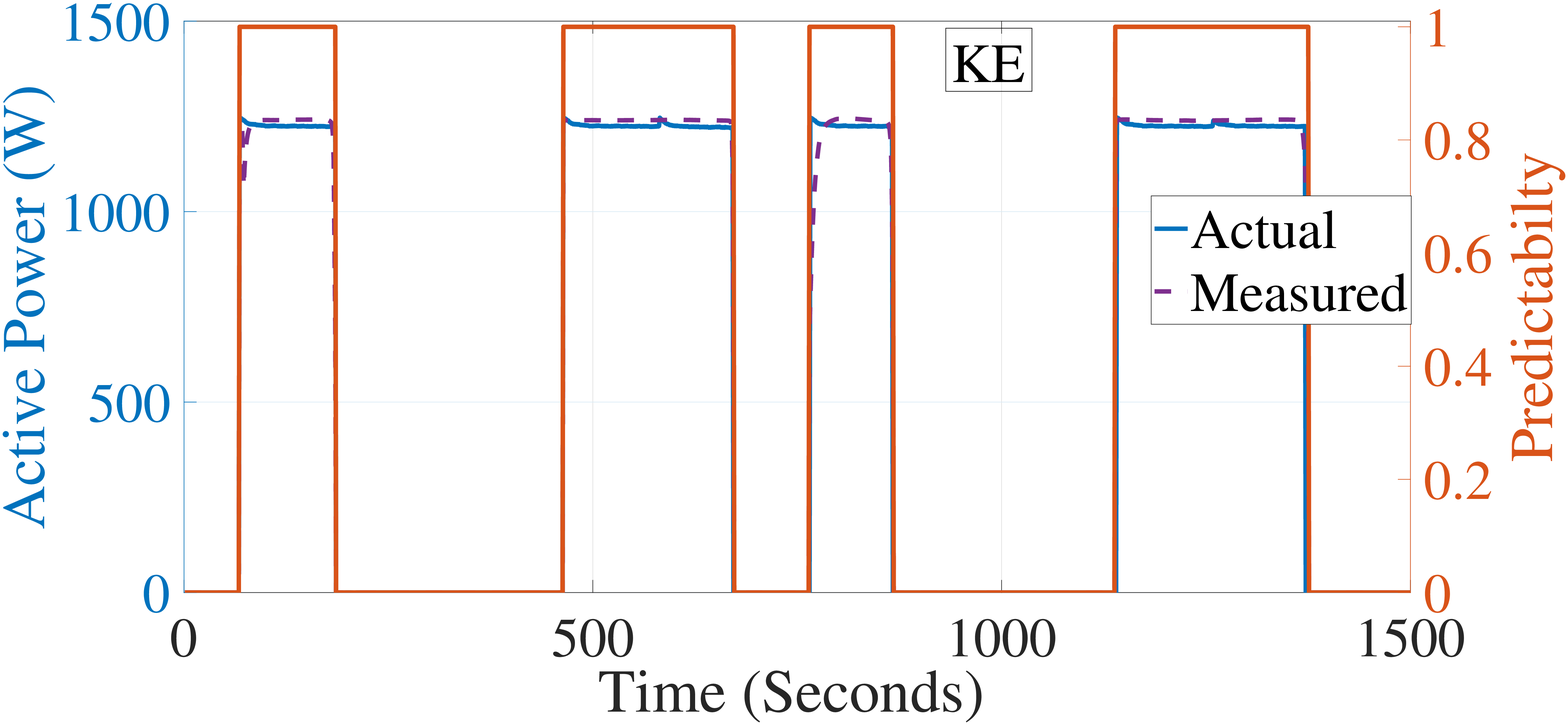}
\subcaption{Load profile of an electric kettle}  \label{c7faapu}
\end{subfigure}
\end{minipage}
\begin{minipage}{0.45\textwidth}
\begin{subfigure}[b]{\textwidth}
\includegraphics[width=\linewidth]{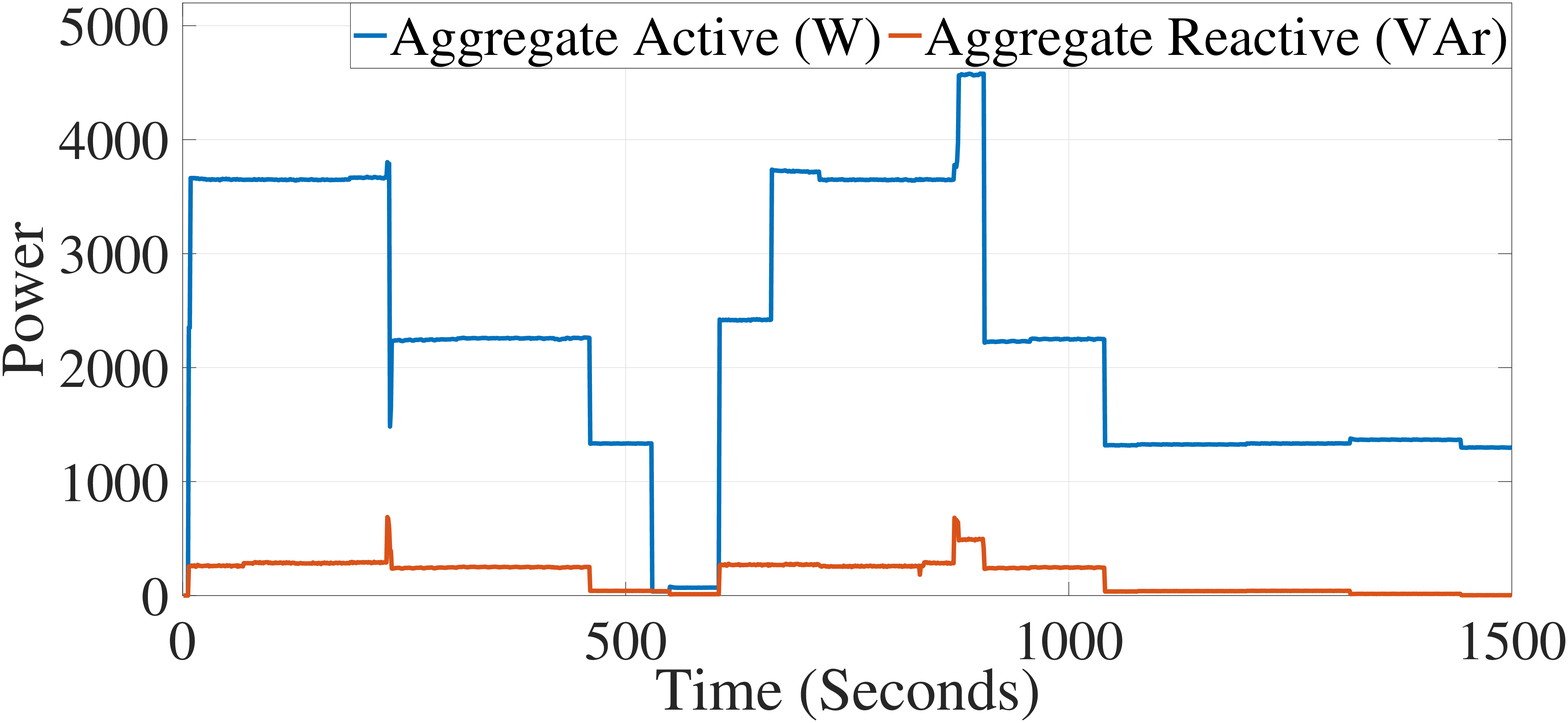}
\end{subfigure}
\begin{subfigure}[b]{\textwidth}
\includegraphics[width=\linewidth]{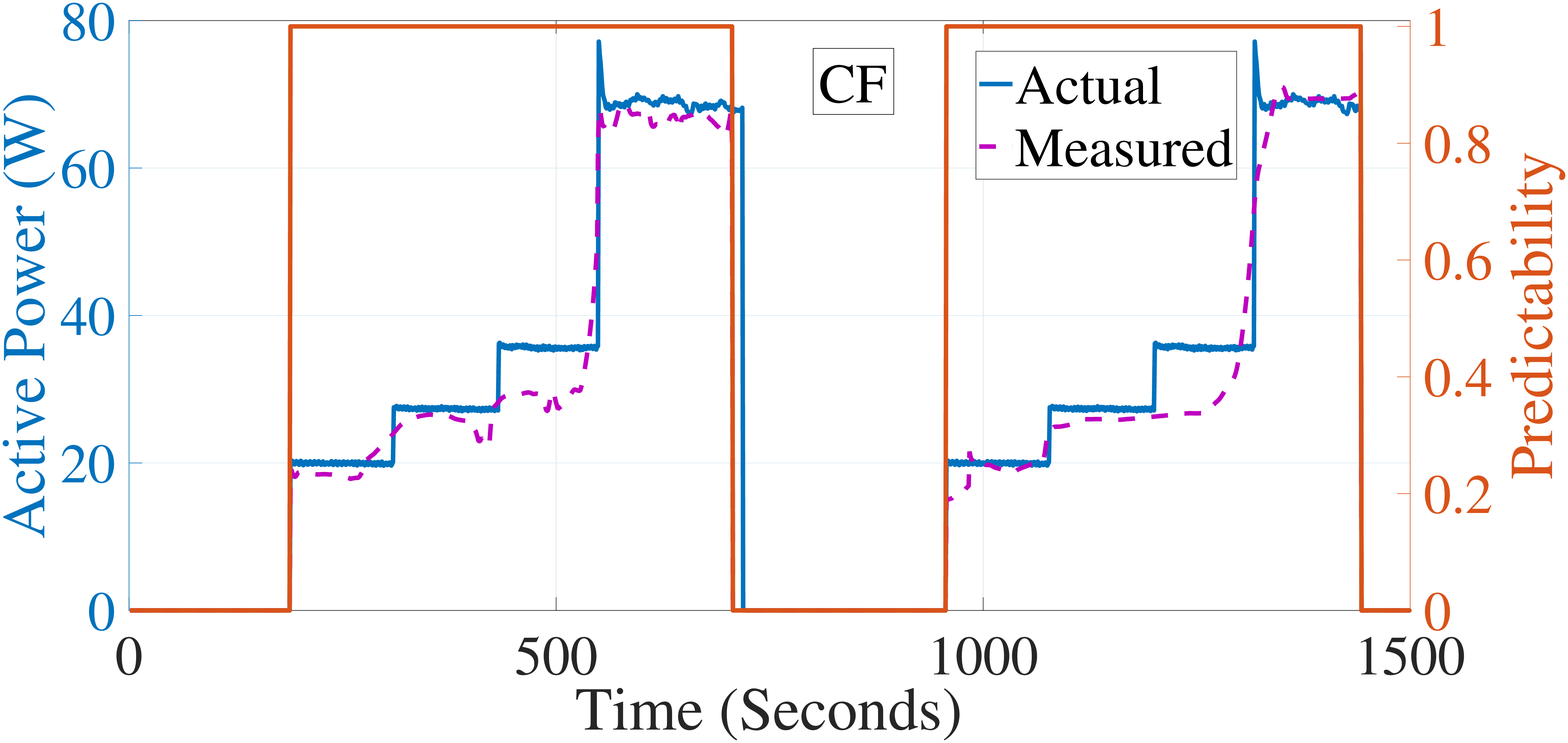}
\subcaption{Load profile of a ceiling fan}  \label{CL8}
\end{subfigure}
\end{minipage}
\noindent\par
\noindent\makebox[\textwidth][c]{%
\begin{minipage}{0.45\textwidth}
\begin{subfigure}[b]{\textwidth}
\includegraphics[width=\linewidth]{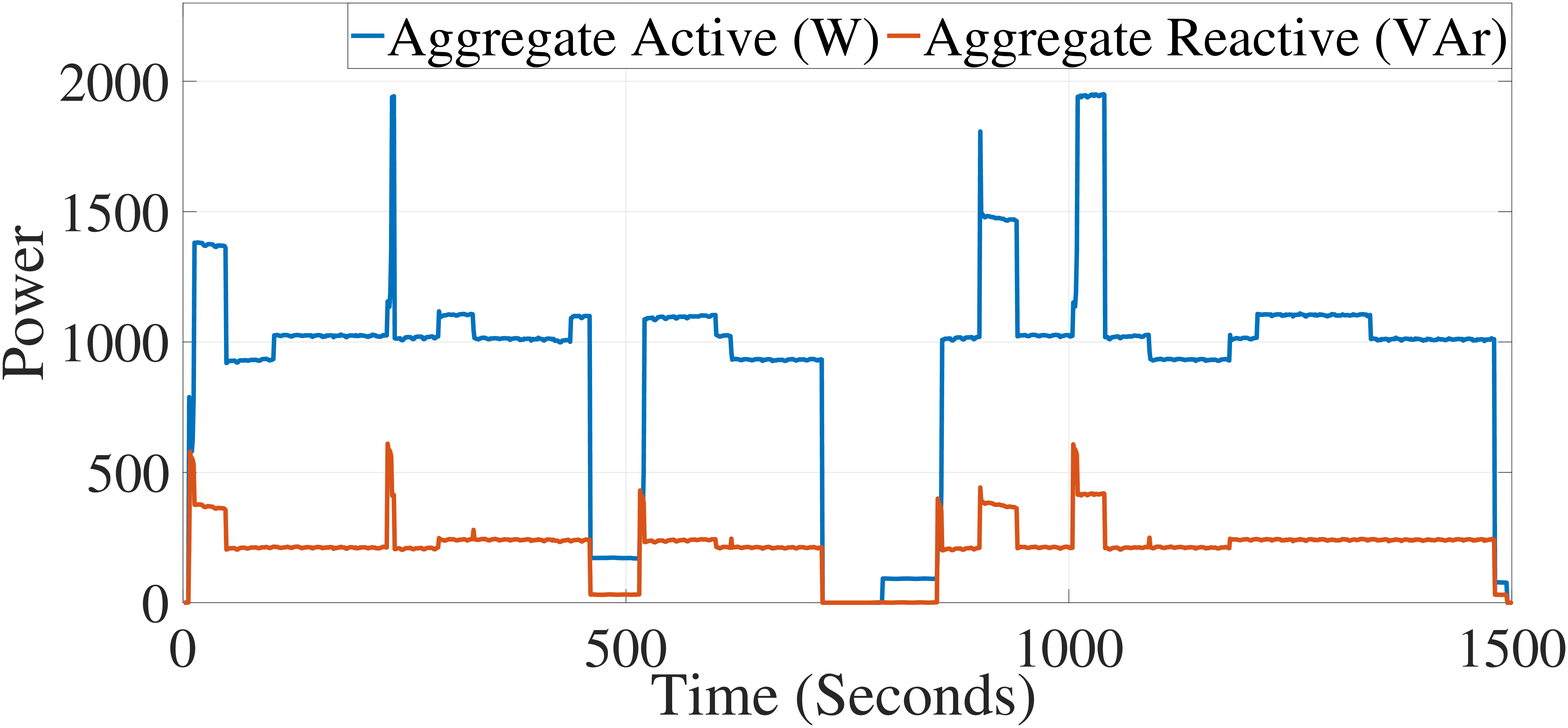}
\end{subfigure}
\begin{subfigure}{\textwidth}
\includegraphics[width=\linewidth]{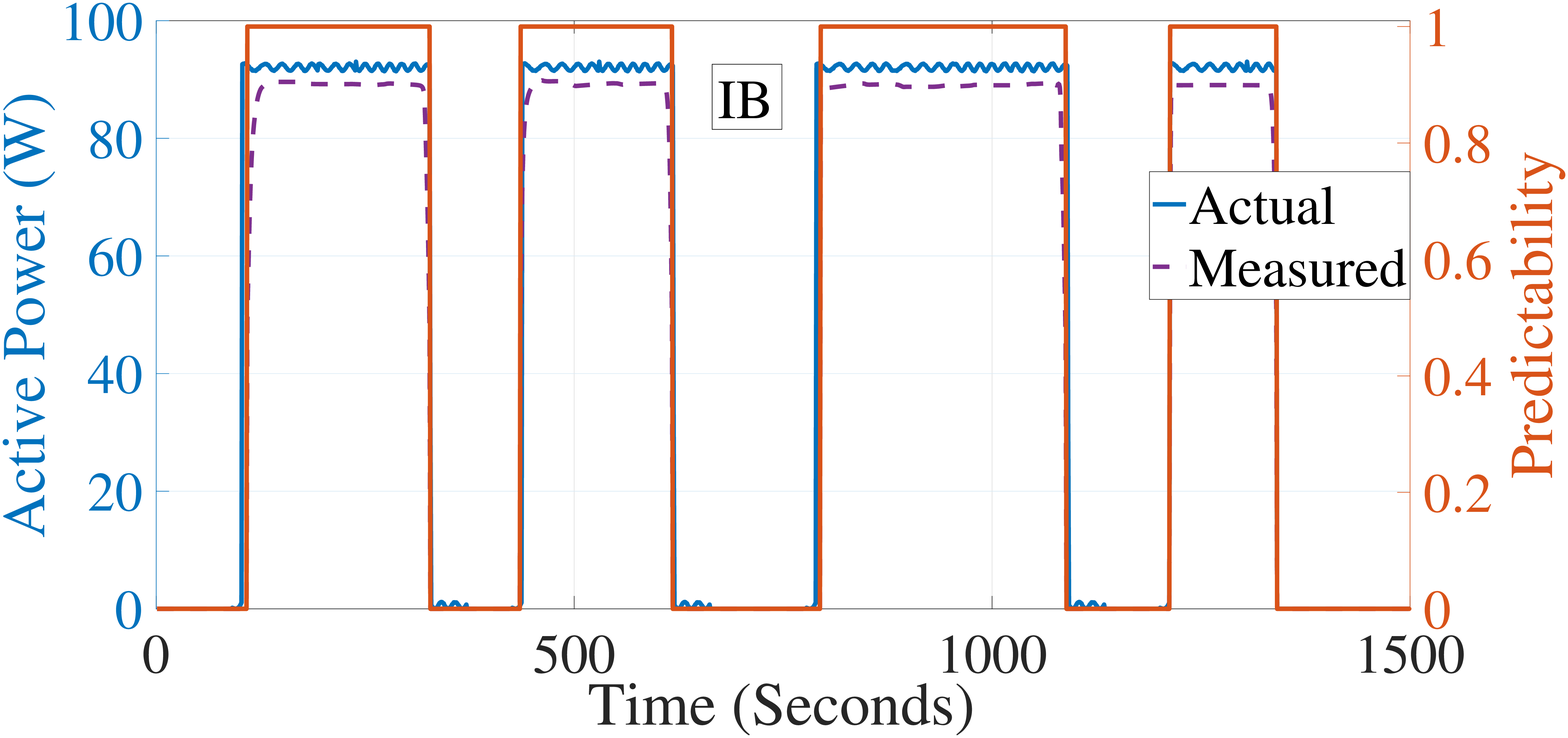}
\subcaption{Load profile of an incandescent bulb}  \label{CL9}
\end{subfigure}
    \hrulefill\par
\end{minipage}}
\caption{Performance of proposed Mh-Net NILM model under constant supply voltage.}
\end{figure*}

\begin{figure*}\centering
\begin{minipage}{0.45\textwidth}
\begin{subfigure}[b]{\textwidth}
\includegraphics[width=\linewidth]{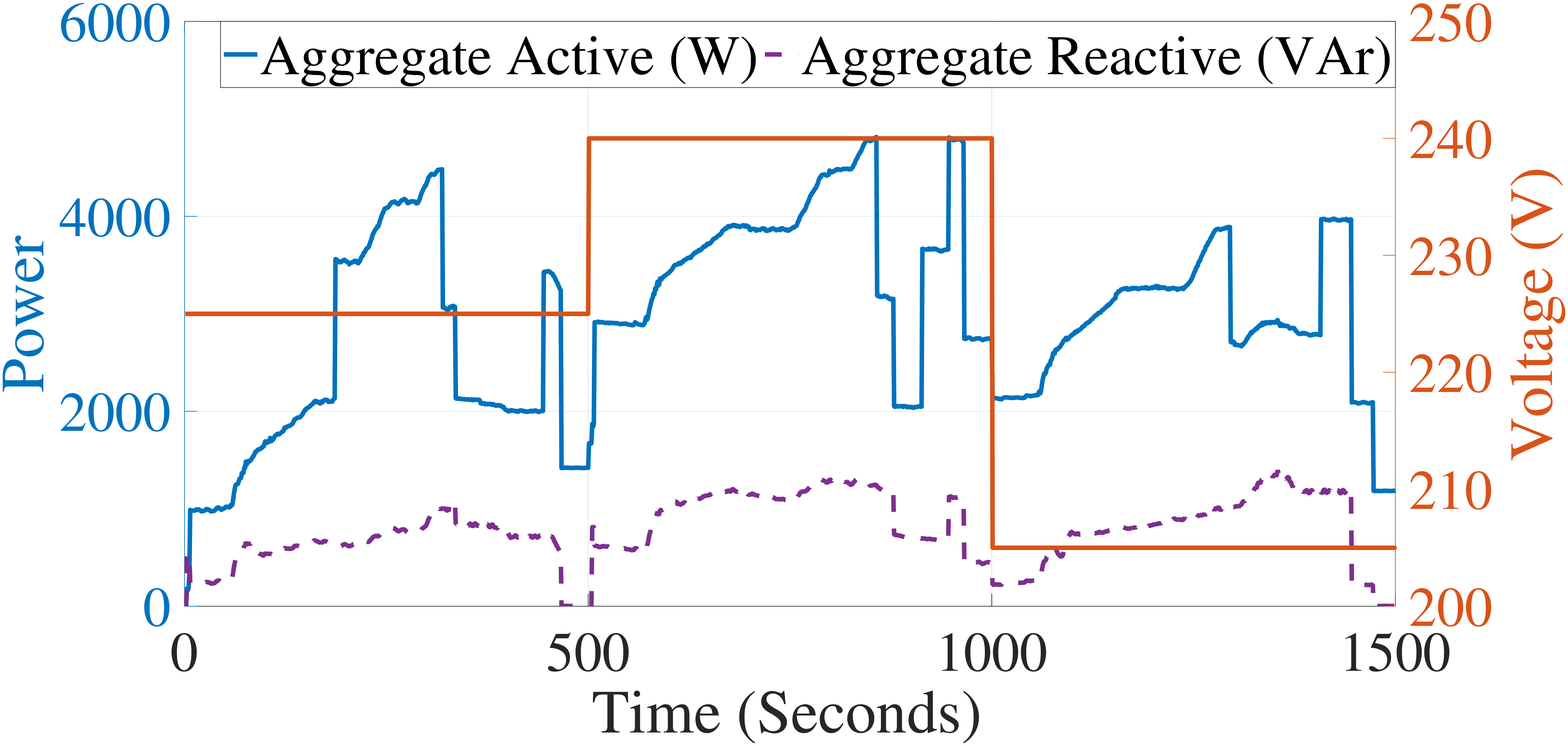}
\end{subfigure}
\begin{subfigure}[b]{\textwidth}
\includegraphics[width=\linewidth]{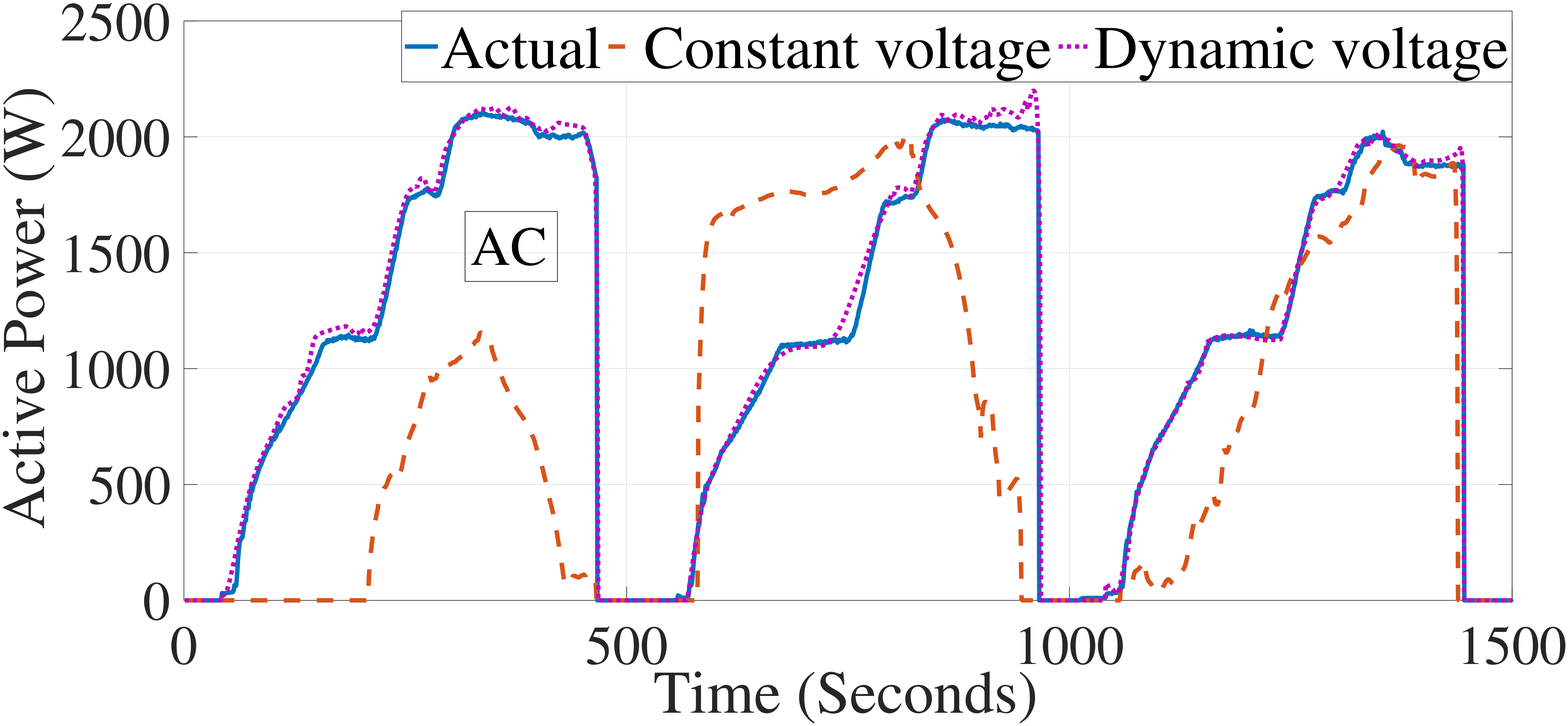}
\subcaption{Load profile of an air-conditioner.}  \label{DL1}
\end{subfigure}
\end{minipage}
\begin{minipage}{0.45\textwidth}
\begin{subfigure}[b]{\textwidth}
\includegraphics[width=\linewidth]{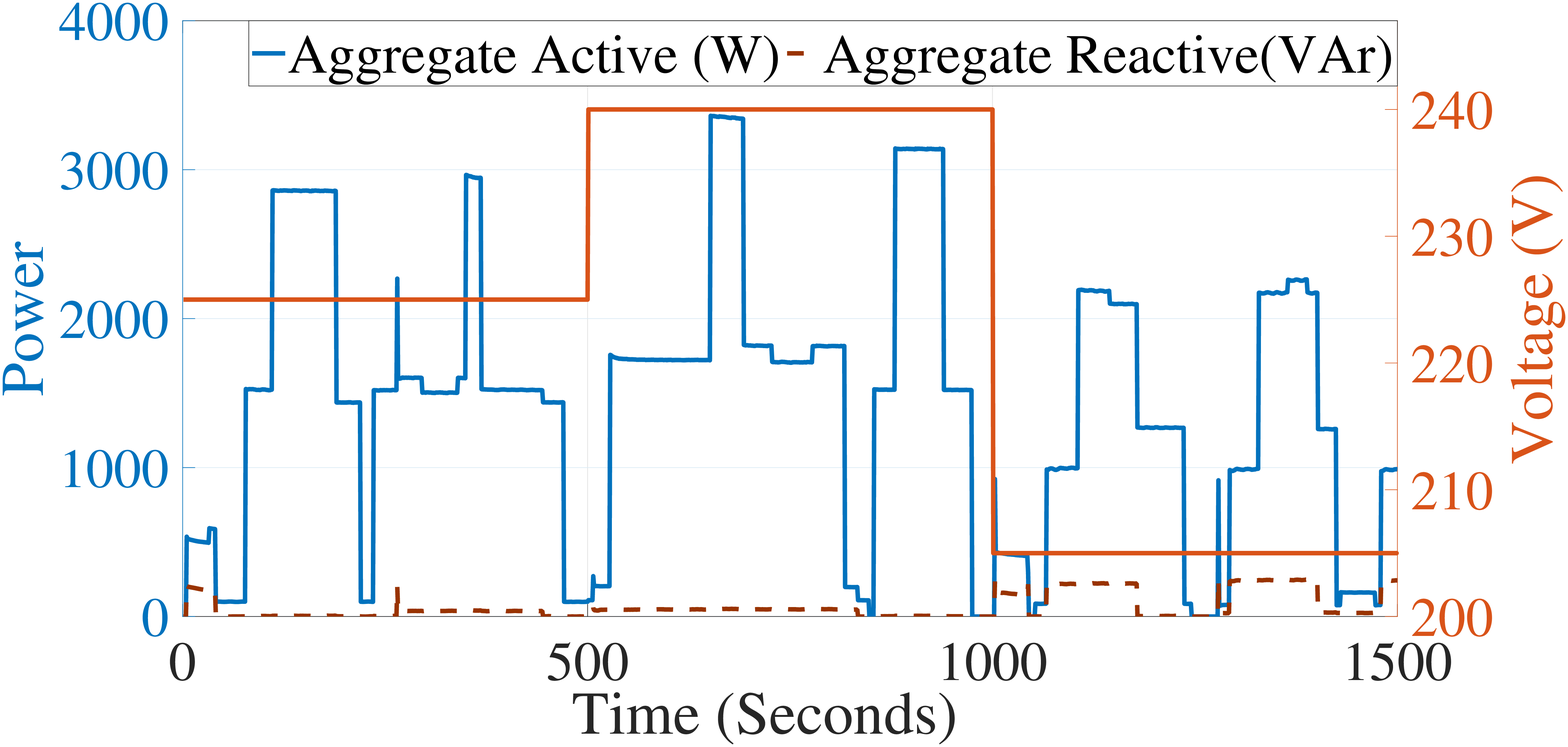}
\end{subfigure}
\begin{subfigure}[b]{\textwidth}
\includegraphics[width=\linewidth]{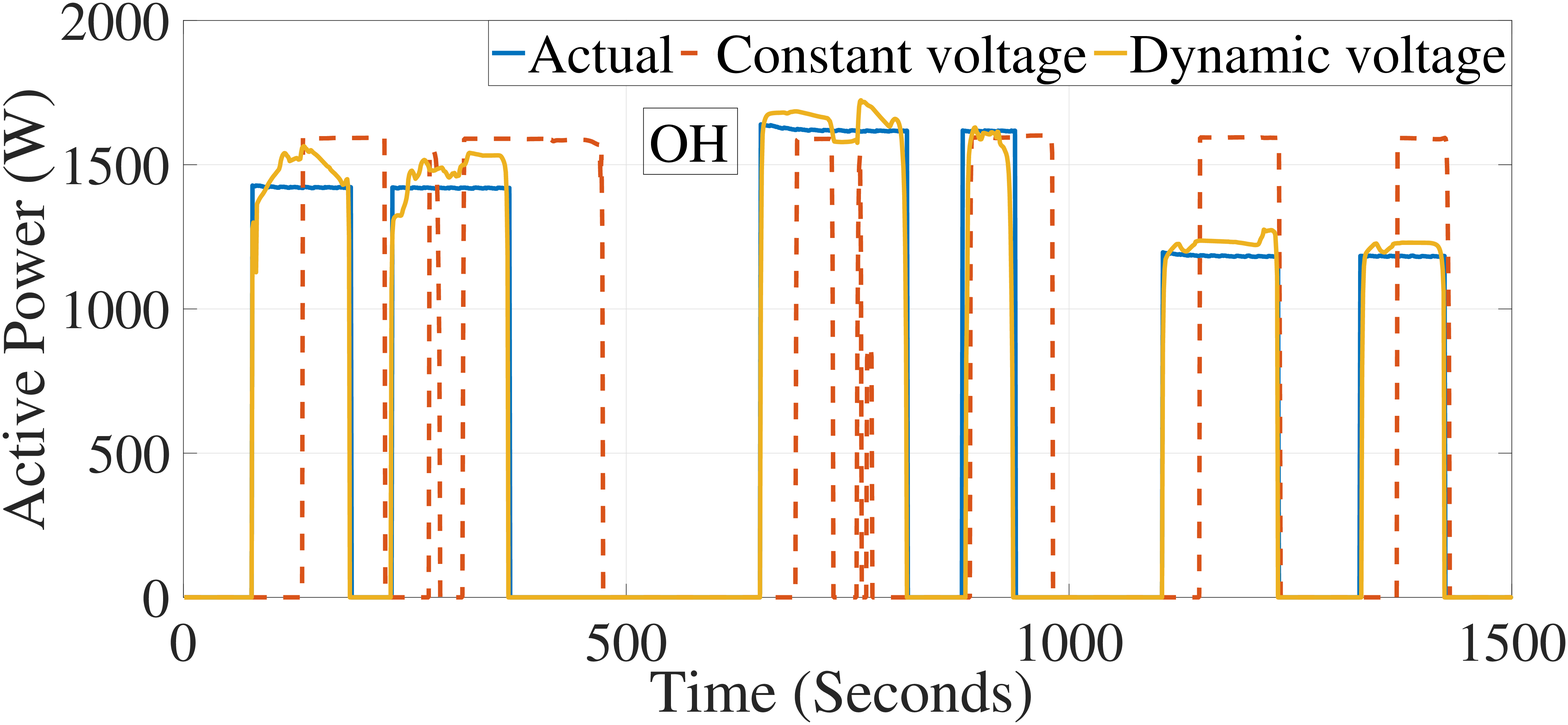}
\subcaption{Load profile of an oil heater.}  \label{DL2}
\end{subfigure}
\end{minipage}
\caption{Performance of proposed Mh-Net NILM model under dynamic supply voltage.}
\end{figure*}

\begin{figure*}\centering
\ContinuedFloat
\begin{minipage}{0.45\textwidth}
\begin{subfigure}[b]{\textwidth}
\includegraphics[width=\linewidth]{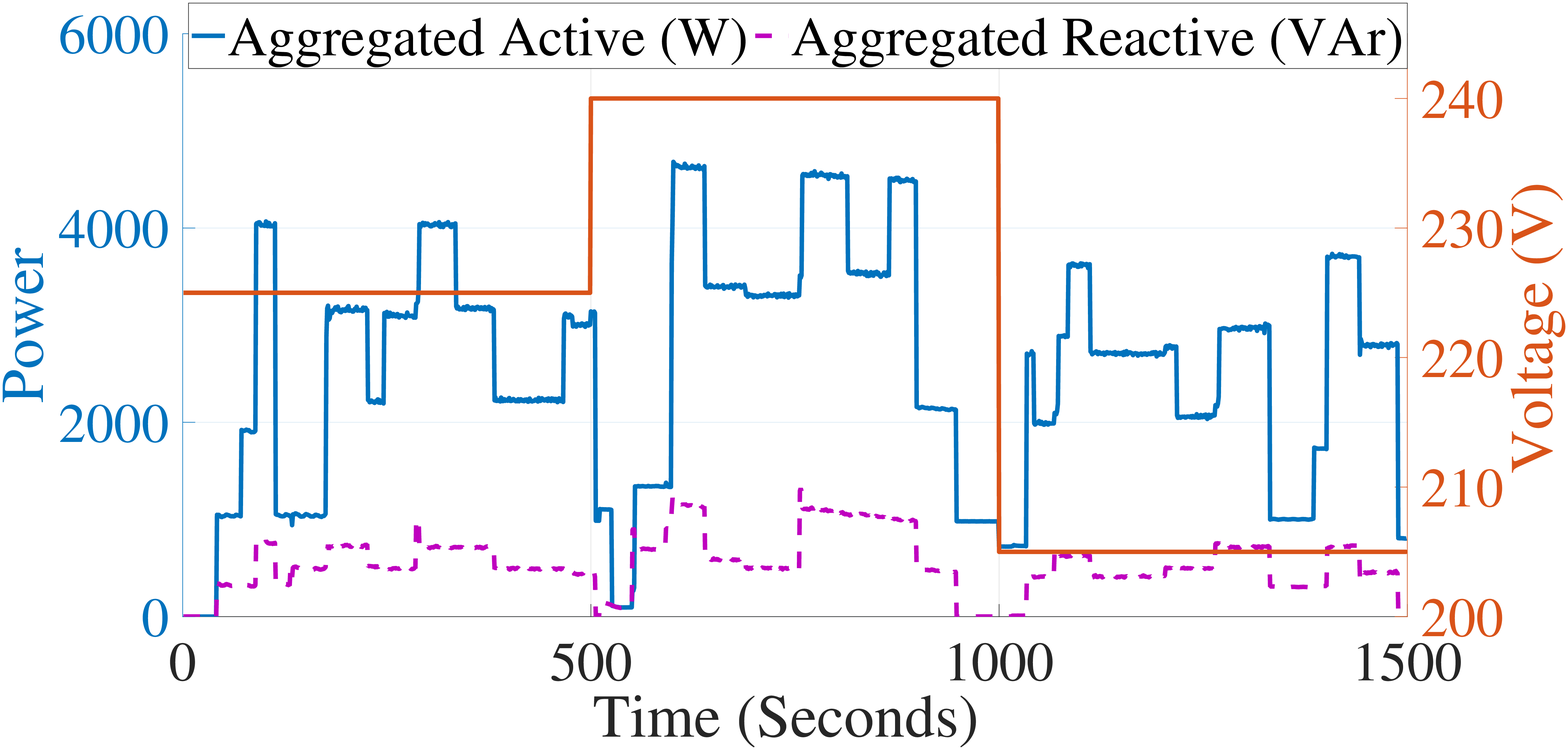}
\end{subfigure}
\begin{subfigure}[b]{\textwidth}
\includegraphics[width=\linewidth]{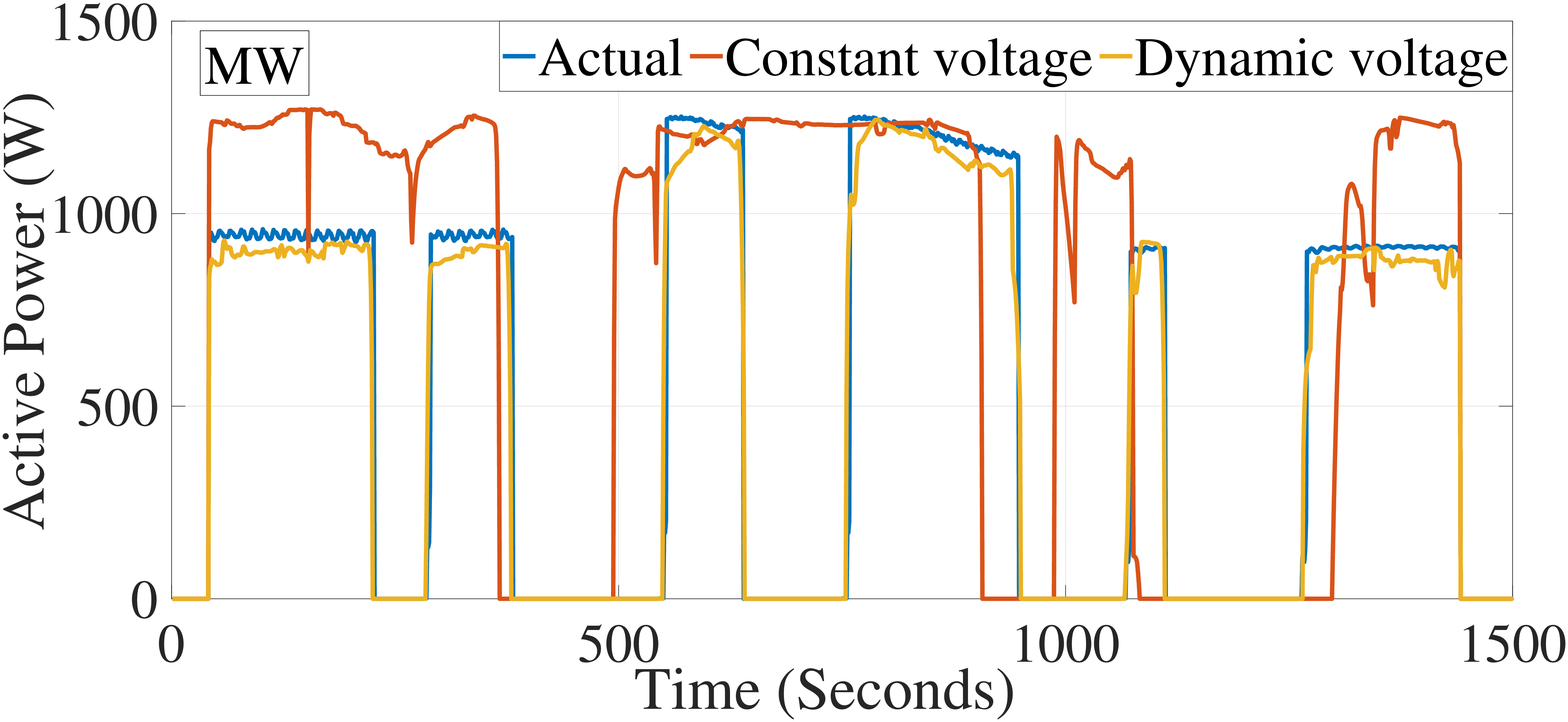}
\subcaption{Load profile of a microwave.}  \label{DL3}
\end{subfigure}
\end{minipage}
\begin{minipage}{0.45\textwidth}
\begin{subfigure}[b]{\textwidth}
\includegraphics[width=\linewidth]{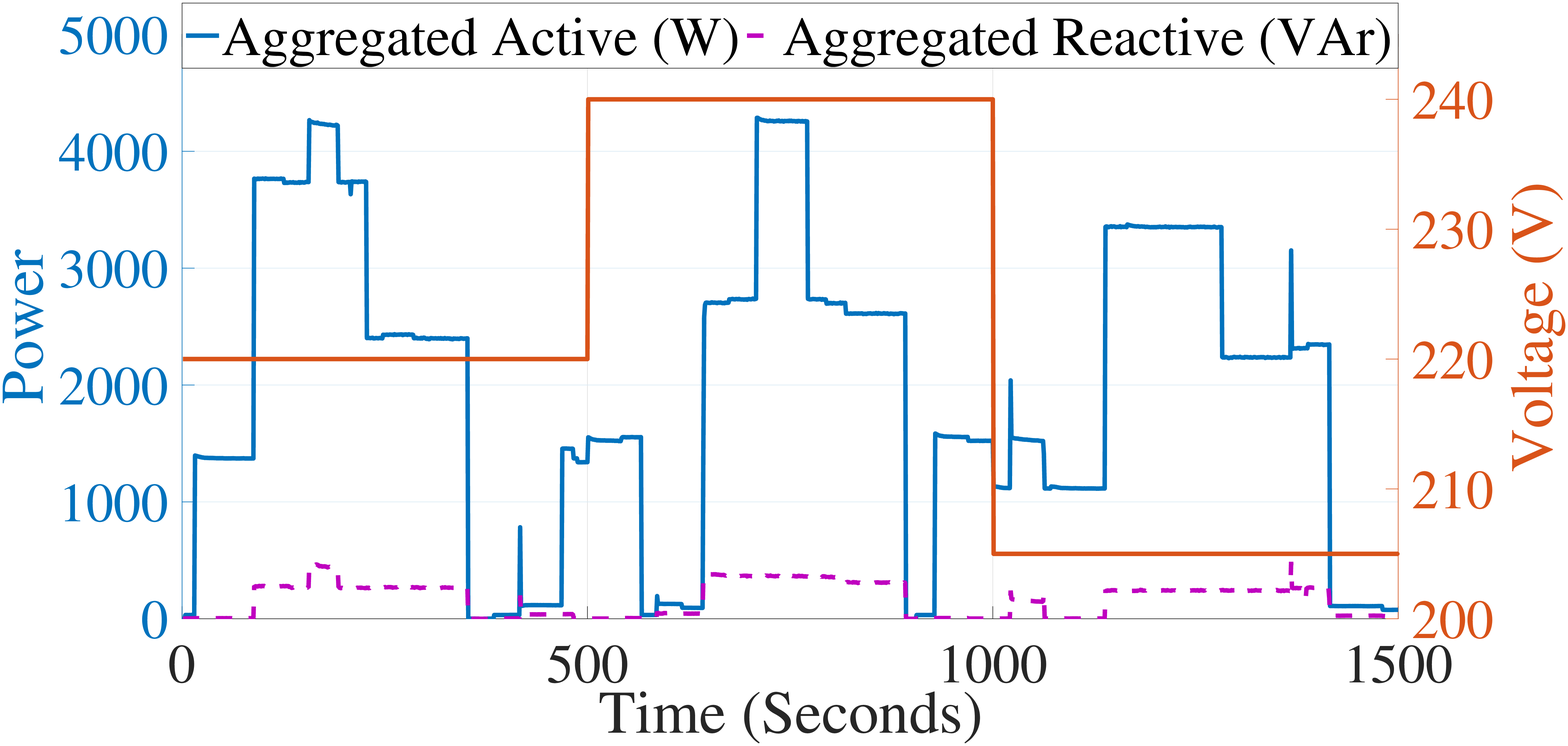}
\end{subfigure}
\begin{subfigure}[b]{\textwidth}
\includegraphics[width=\linewidth]{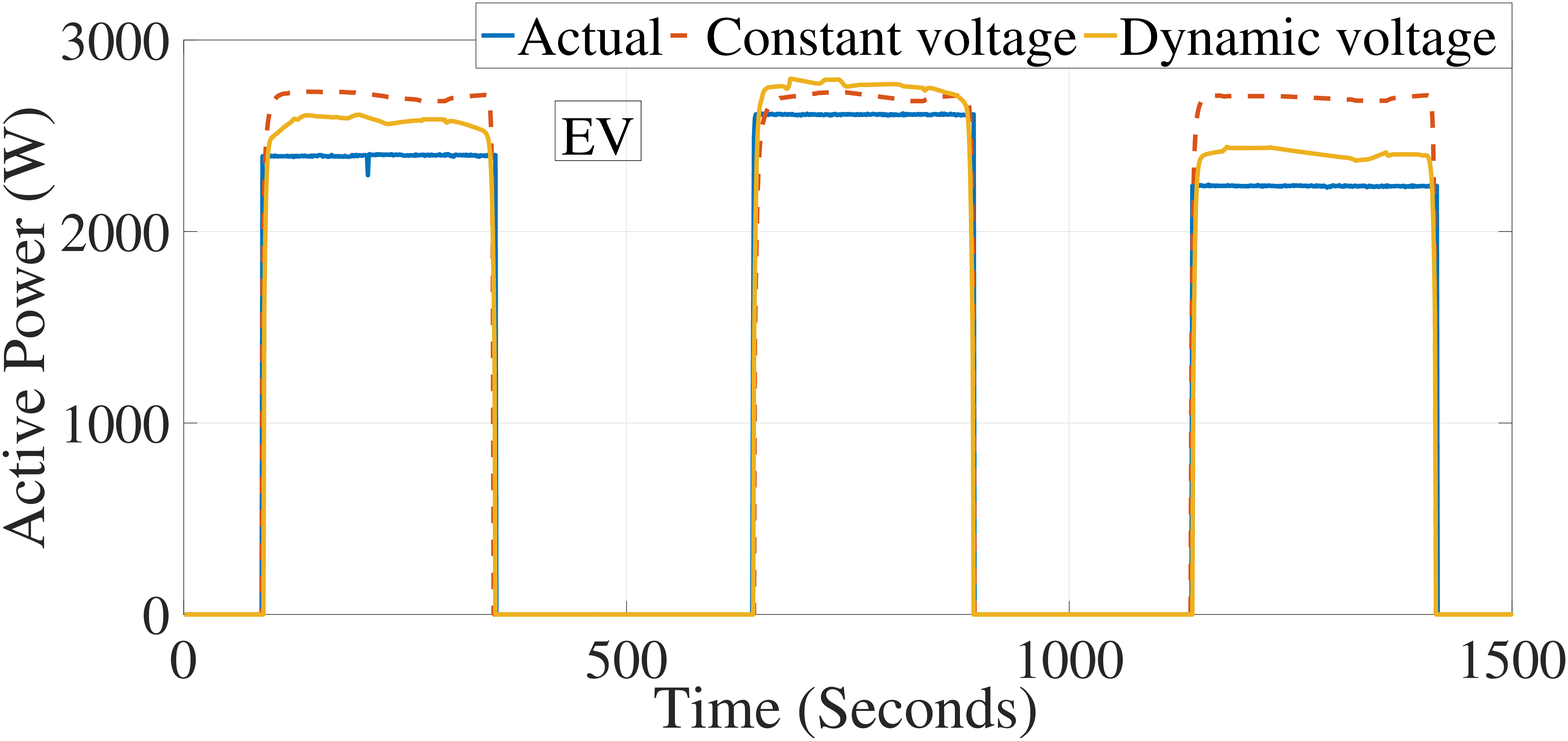}
\subcaption{Load profile of an electric vehicle.}  \label{DL4}
\end{subfigure}
\end{minipage}
\begin{minipage}{0.45\textwidth}
\begin{subfigure}[b]{\textwidth}
\includegraphics[width=\linewidth]{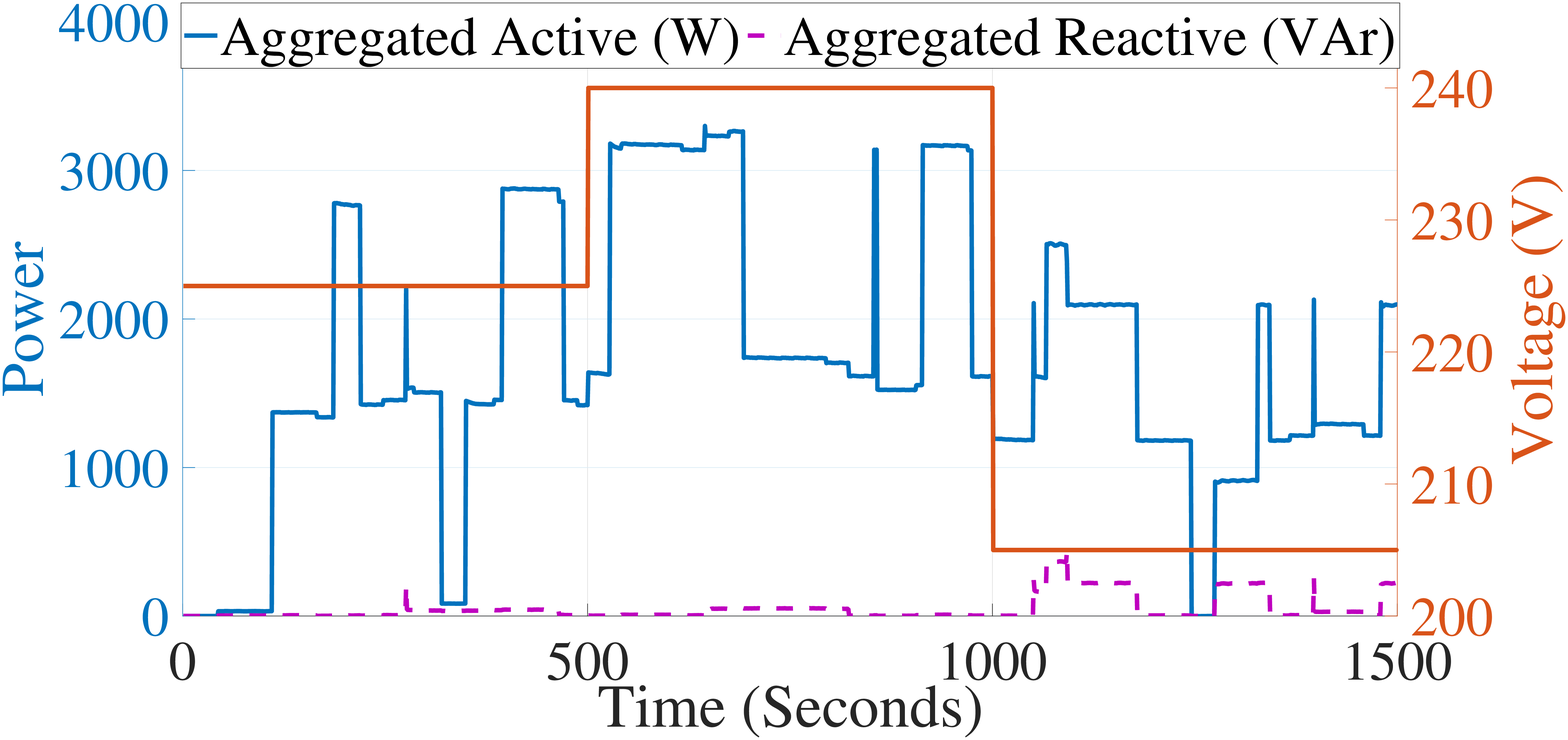}
\end{subfigure}
\begin{subfigure}[b]{\textwidth}
\includegraphics[width=\linewidth]{CaseC_L52.eps}
\subcaption{Load profile of an refrigerator.}  \label{DL5}
\end{subfigure}
\end{minipage}
\begin{minipage}{0.45\textwidth}
\begin{subfigure}[b]{\textwidth}
\includegraphics[width=\linewidth]{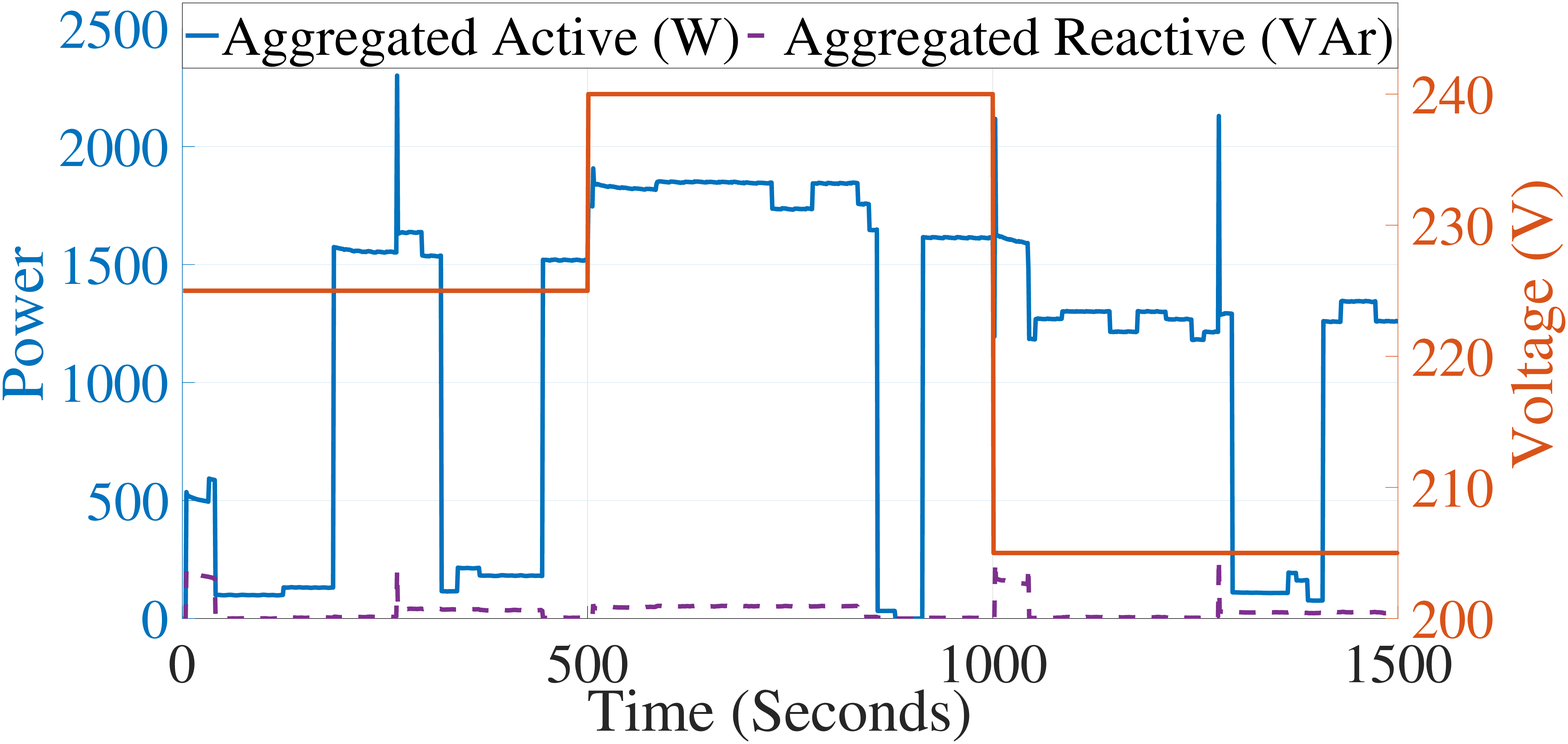}
\end{subfigure}
\begin{subfigure}[b]{\textwidth}
\includegraphics[width=\linewidth]{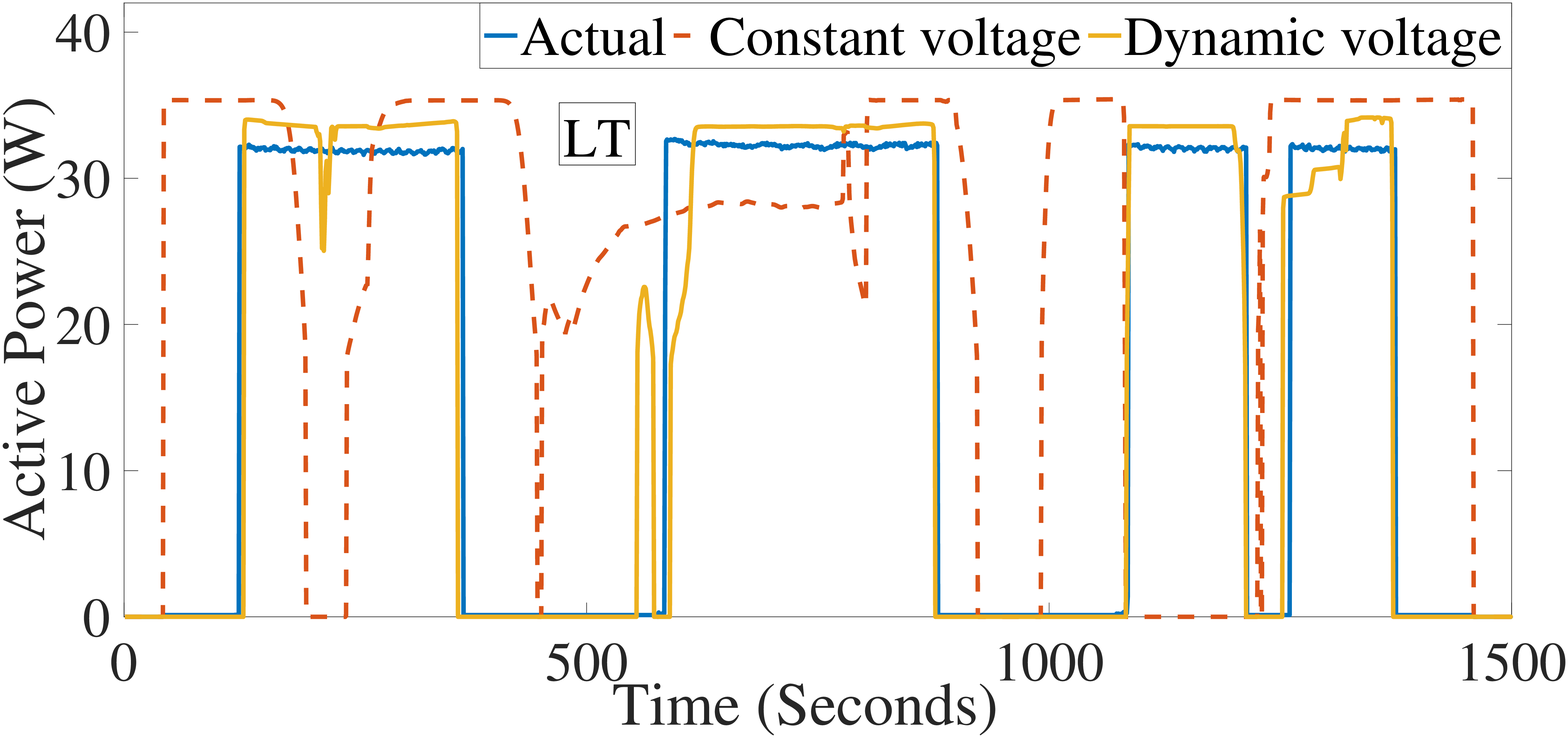}
\subcaption{Load profile of a LED tube.}  \label{DL6}
\end{subfigure}
\end{minipage}
\begin{minipage}{0.45\textwidth}
\begin{subfigure}[b]{\textwidth}
\includegraphics[width=\linewidth]{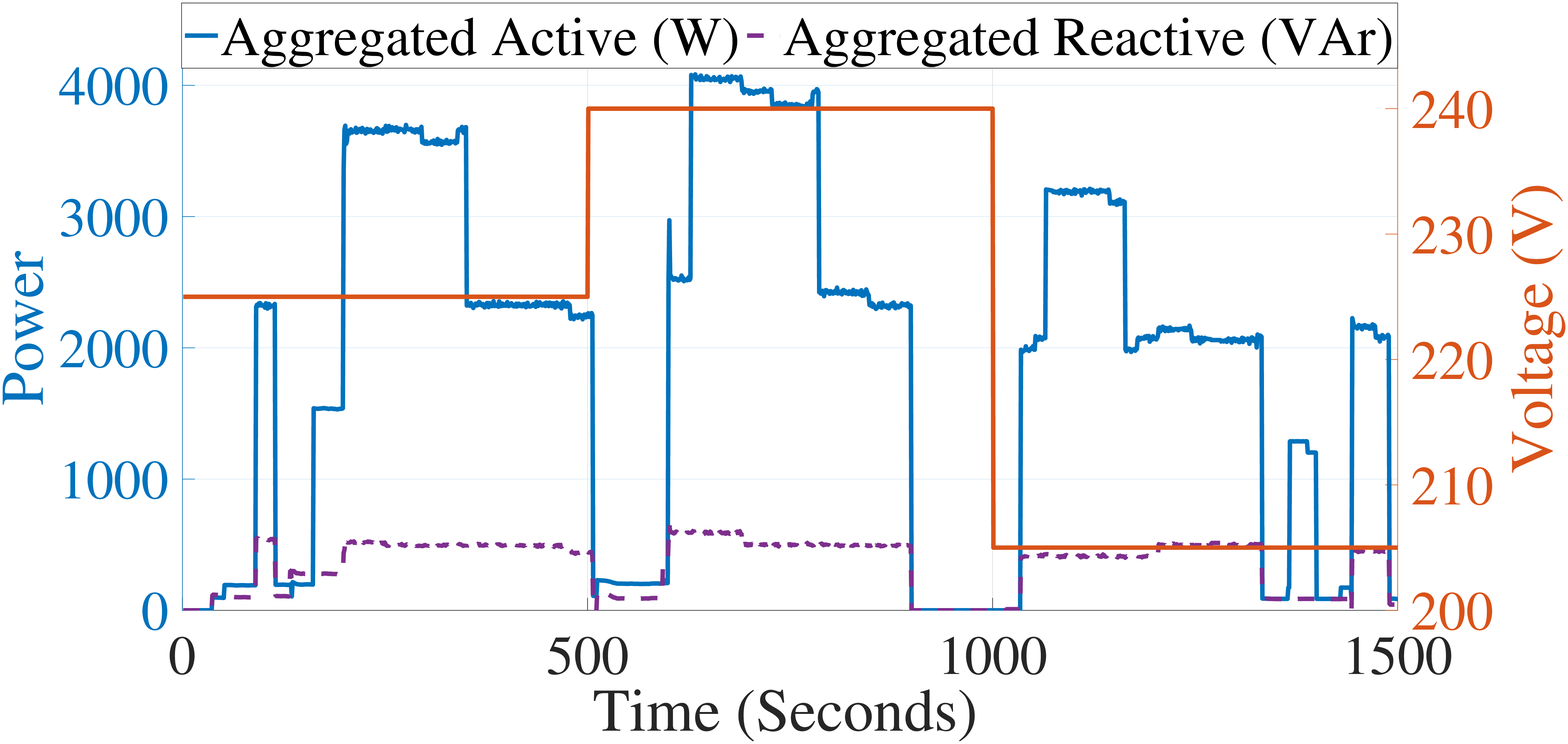}
\end{subfigure}
\begin{subfigure}[b]{\textwidth}
\includegraphics[width=\linewidth]{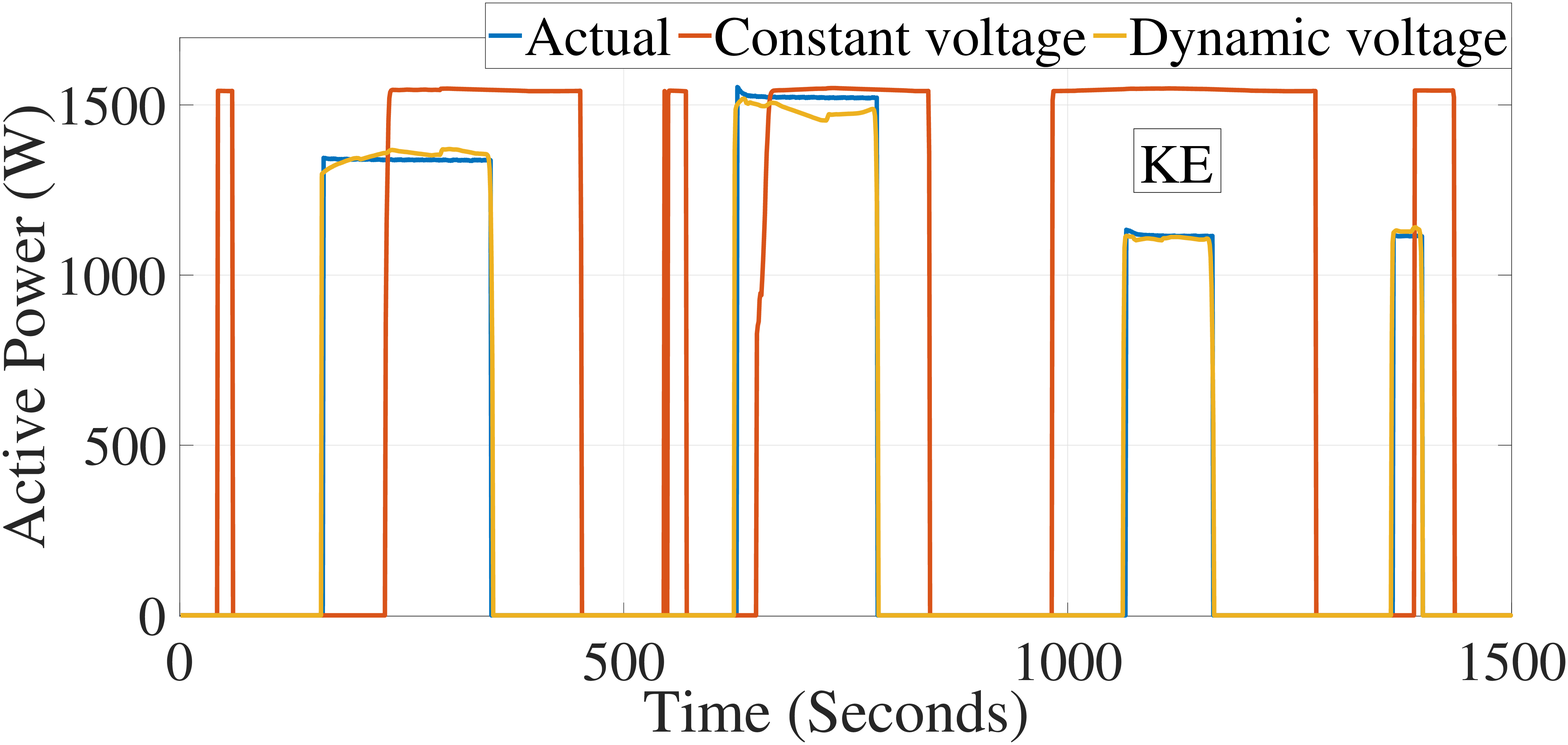}
\subcaption{Load profile of an electric kettle.}  \label{DL7}
\end{subfigure}
\end{minipage}
\begin{minipage}{0.45\textwidth}
\begin{subfigure}[b]{\textwidth}
\includegraphics[width=\linewidth]{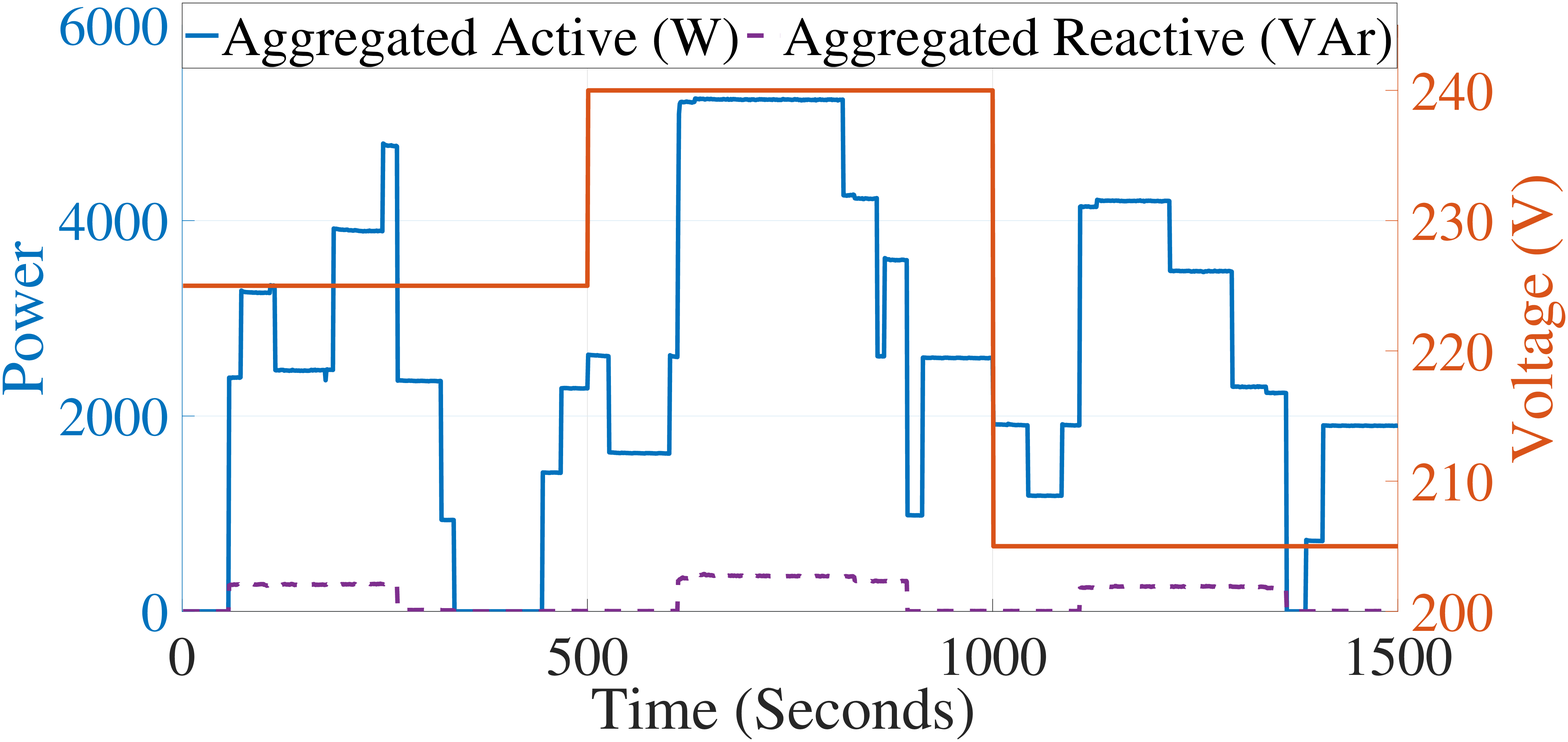}
\end{subfigure}
\begin{subfigure}[b]{\textwidth}
\includegraphics[width=\linewidth]{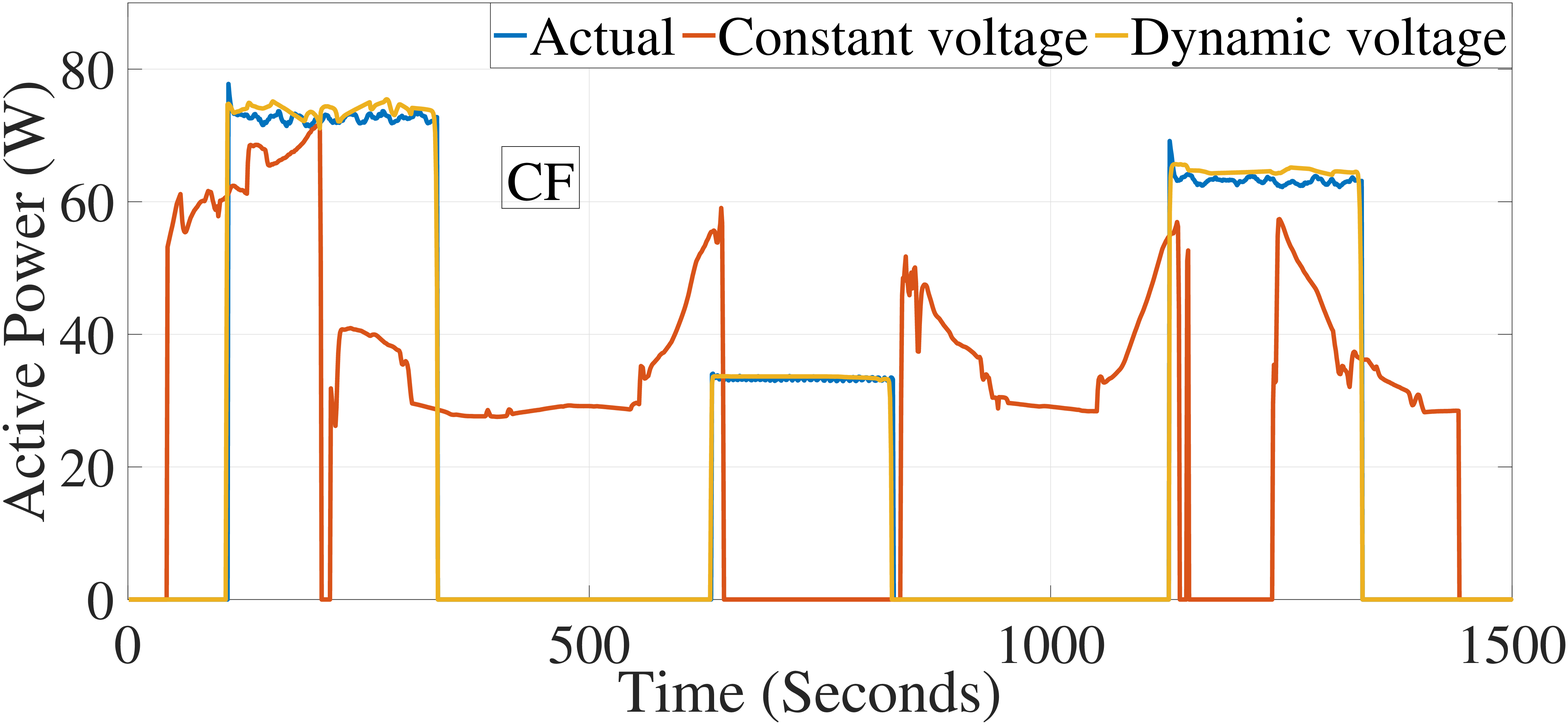}
\subcaption{Load profile of a ceiling fan.}  \label{DL8}
\end{subfigure}
\end{minipage}
\caption{Performance of proposed Mh-Net NILM model under dynamic supply voltage.}
\end{figure*}
\begin{figure*}\centering
\ContinuedFloat
\begin{minipage}{0.45\textwidth}
\begin{subfigure}[b]{\textwidth}
\includegraphics[width=\linewidth]{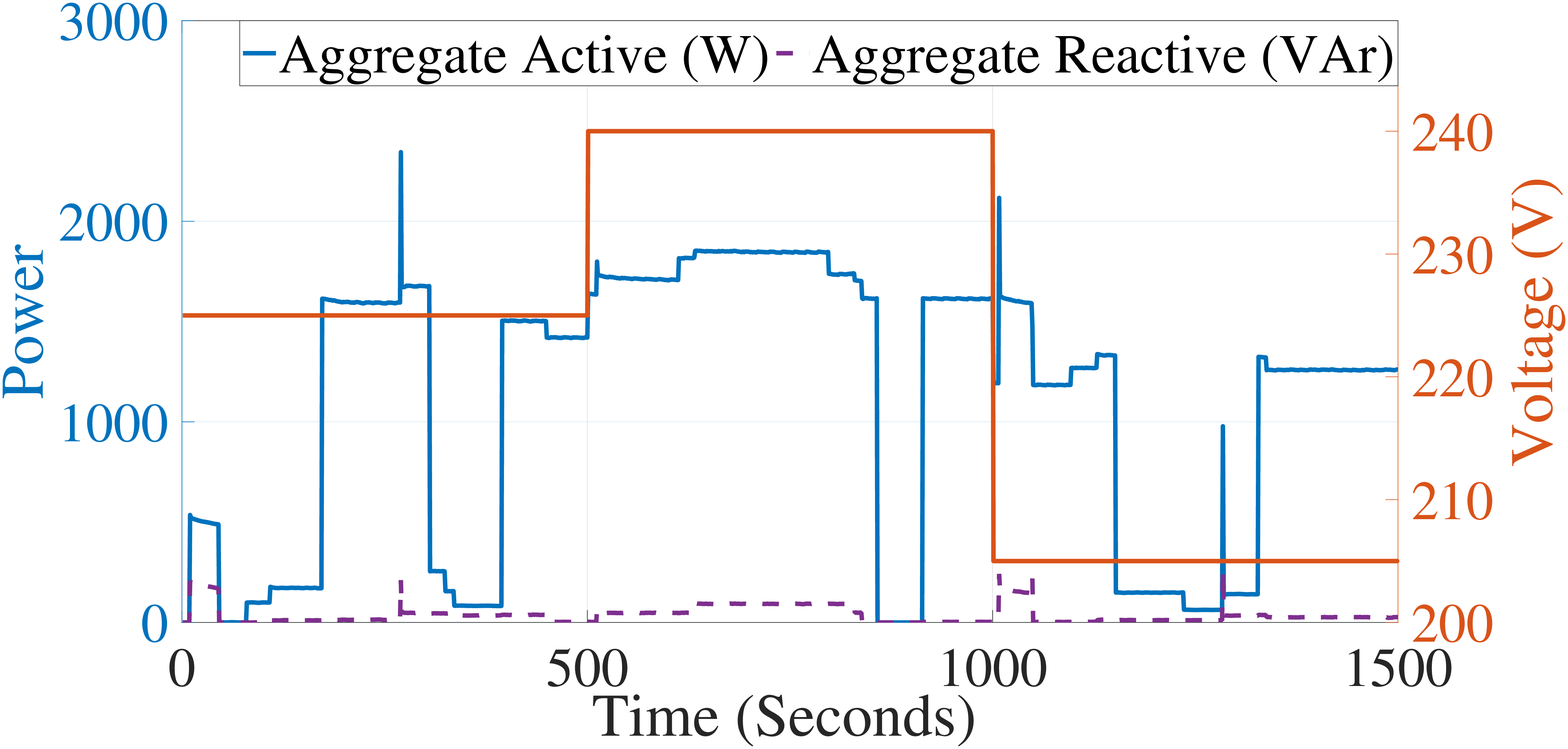}
\end{subfigure}
\begin{subfigure}[b]{\textwidth}
\includegraphics[width=\linewidth]{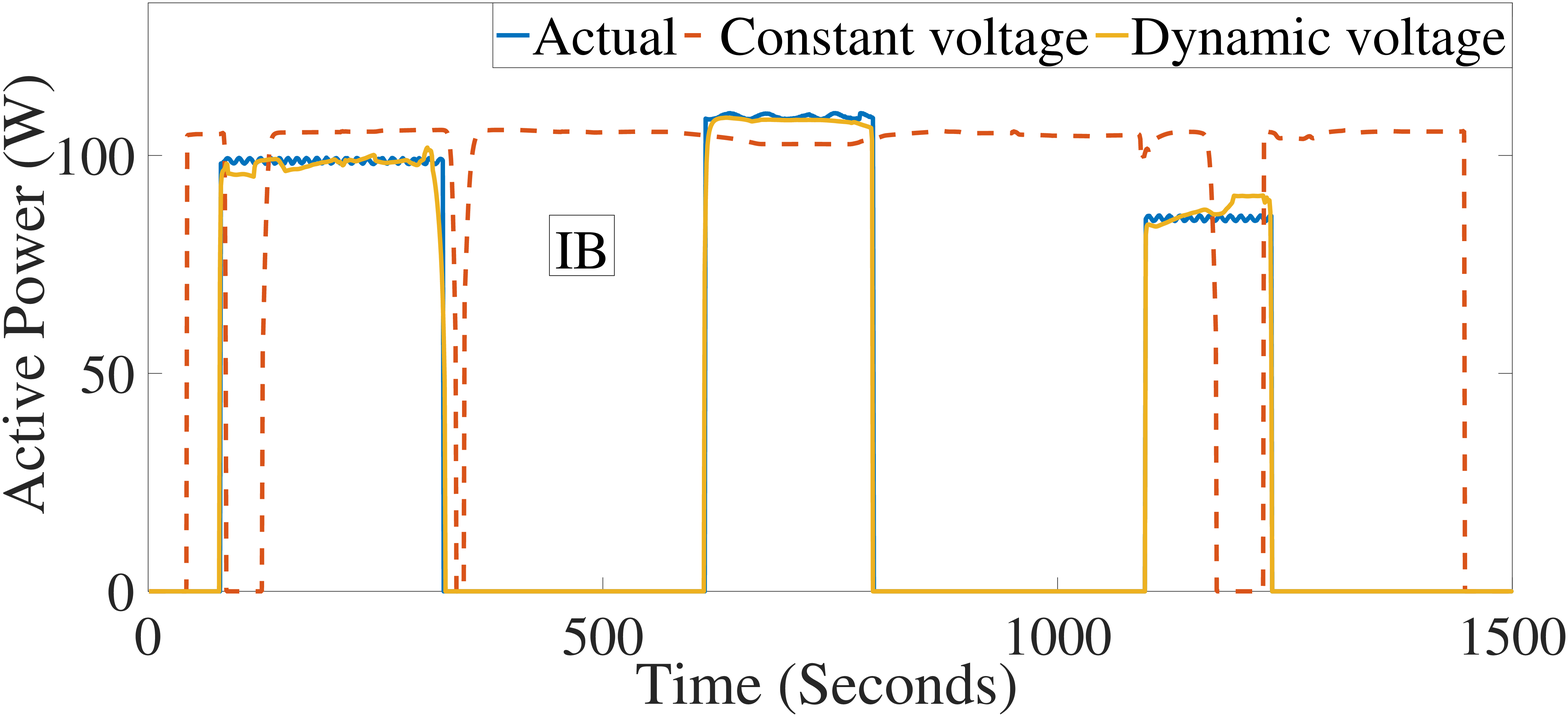}
\subcaption{Load profile of an incandescent bulb.}  \label{DL9}
\end{subfigure}
\end{minipage}
\caption{Performance of proposed Mh-Net NILM model under dynamic supply voltage.}
\end{figure*}
\begin{table}[h]
\caption{Performance metrics under constant voltage.}\label{step}\centering
    \begin{tabular}{P{0.75cm}|P{0.6cm}|P{0.6cm}|P{0.6cm}|P{0.6cm}|P{0.6cm}|P{0.6cm}|P{0.6cm}|P{0.6cm}|P{0.6cm}}
        \hline
        \hline
        ~ & AC & OH & MW & EV & RF & LT & KE & CF & IB \\
        \hline
        MAE & 14.34 & 59.39 & 27.22 & 85.19 & 25.42 & 3.04 & 14.59 & 2.60 & 3.19 \\
        \hline
        SAE & 4.88 & 3.69 & 2.44 & 3.31 & 1.37 & 0.65 & 7.57 & 2.04 & 3.04\\
        \hline
        F1(\%) & 86.06 & 98.61 & 99.36 & 99.32 & 96.30 & 85.32 & 99.69 & 99.59 & 99.36\\
         \hline
        \hline
    \end{tabular}
    \vspace{-0.6cm}
\end{table}
\begin{table}[h]
\caption{Performance metrics under dynamic voltage conditions.}\label{variable}\centering
    \begin{tabular}{P{1.0cm}|P{1.25cm}|P{1.25cm}|P{1.1cm}|P{1.25cm}|P{1.25cm}|P{1.1cm}}
        \hline
        \hline
        ~ & \multicolumn{3}{c|}{with constant voltage training} & \multicolumn{3}{c}{with dynamic voltage training} \\
        \hline
        ~ & MAE & SAE & F1 (\%) ~ & MAE & SAE & F1 (\%) \\
        \hline
        \rule{0pt}{2ex} AC & 579.02 & 355.93 & 90.63 & 30.05 & 22.37 & 98.78\\
        \hline
        \rule{0pt}{2ex} OH & 578.06 & 70.24 & 78.86 & 46.13 & 2.31 & 99.43 \\
        \hline
        \rule{0pt}{2ex} MW & 410.54 & 236.14 & 81.42 & 35.87 & 20 & 99.56\\
        \hline
        \rule{0pt}{2ex} EV & 166.53 & 133.70 & 99.73 & 103.25 & 61.90 & 99.73\\
        \hline
        \rule{0pt}{2ex} RF & 65.29 & 8.36 & 68.36 & 6.55 & 6.10 & 99.80\\
        \hline
        \rule{0pt}{2ex} LT & 16.75 & 7.62 & 68.53 & 1.95 & 0.22 & 98.23\\
        \hline
        \rule{0pt}{2ex} KE  & 602.81 & 387.96 & 77.50 & 18.33 & 6.40 & 99.53\\
        \hline
        \rule{0pt}{2ex} CF & 31.67 & 5.05 & 45.99 & 0.65 & 0.38 & 99.83 \\
        \hline
        \rule{0pt}{2ex} IB & 66.09 & 53.14 & 55.83 & 0.91 & 0.11 & 99.83 \\
         \hline
        \hline
    \end{tabular}
    \vspace{-0.6cm}
\end{table}
\subsection{Performance evaluation of proposed Mh-Net CNN NILM model under dynamic voltage grid scenario.}
To evaluate the performance of the proposed NILM technique in practical grid conditions, the Mh-Net CNN model trained at constant voltage is provided with aggregated active and reactive power datasets under dynamic voltage conditions. The performance of this trained model under constant voltage dataset is evaluated, and is observed to demonstrate deteriorating performance under dynamic supply voltage. In order to ensure robust performance under practical grid operating conditions, the proposed Mh-Net CNN model has been trained under dynamic grid voltage conditions, which are representative of practical grid operation. Similar to the previous case study, nine test scenarios have been considered and the supply voltage is varied within the acceptable range, i.e. 205V to 240V @ 50Hz. A comparison of performance of the proposed Mh-Net CNN model, with constant voltage training and dynamic voltage training is shown in Figs. \ref{DL1}-\ref{DL2}. It is observed from the figures that under different voltage scenarios, the performance of Mh-Net CNN model trained with constant voltage dataset degrades drastically. However, the proposed Mh-Net CNN model trained with the dynamic voltage datasets demonstrates superior performance under dynamic supply voltage conditions. \\
\indent A comparison of the performance metrics of the two models (trained with constant and dynamic voltage datasets), is shown in Table \ref{variable}. It can be observed from the performance matrices that the values of MAE, SAE and F1 score for OH are improved from 578.06, 70.24 and 78.86 for constant voltage-trained model to 46.13, 2.31 and 99.43 for dynamic voltage-trained model respectively. Similarly, major improvement has been observed in the performance metrics for other appliances. Hence, it can be concluded that the proposed model with dynamic dataset training is accurate and demonstrates robust performance. 
\subsection{Performance evaluation of proposed Mh-Net CNN model on UK-DALE dataset.}
Performance of the proposed Mh-Net CNN model is further evaluated using the real-world UK Domestic Appliance-Level Electric (UK-DALE) dataset \cite{UK-DALE}. An individual and aggregated dataset for five appliances at a sampling frequency $\leq$ 1 Hz has been considered for this case. The considered appliances include dishwasher (DW), refrigerator (RF), electric kettle (KE), microwave (MW) and washing machine (WM). \\
\indent A comparison of the performance metrics of the proposed Mh-Net CNN model with well established NILM techniques tested on the UK-DALE dataset is shown in Table \ref{Ukdale}. It can be observed from the Table \ref{Ukdale} that the average values of MAE and SAE for the proposed approach are obtained as 8.71 and 5.00, respectively which are found to be minimum in comparison to the other approaches. Furthermore, average value of F1 score for the proposed approach is 74.38, which is observed to be maximum as compared to other existing approaches. 
Similar trends have been observed for MAE, SAE and F1 scores of individual appliances, thus, highlighting superior performance of the proposed model in comparison to existing approaches.
\begin{table}
\centering
\caption{Performance of proposed Mh-Net CNN model on UK-DALE dataset}
\label{Ukdale}
\scalebox{0.85}{
\begin{tabular}{llllllll}
\hline
\hline
Model              & Metric  & DW    & RF    & KE    & MW    & WM    & Average \\ \hline
\hline
                   & MAE     & 48.25 & 60.93 & 38.02 & 43.63 & 67.91 & 51.75   \\
FHMM \cite{inproceedings}      & SAE     & 46.04 & 51.9  & 35.41 & 41.52 & 64.15 & 47.8    \\
                   & F1 (\%) & 11.79 & 33.52 & 9.35  & 3.44  & 4.1   & 12.44   \\ \hline
                   & MAE     & 22.18 & 17.72 & 10.87 & 12.87 & 13.64 & 15.46   \\
DAE \cite{BONFIGLI20181461}       & SAE     & 18.24 & 8.74  & 7.95  & 9.99  & 10.67 & 11.12   \\
                   & F1 (\%) & 54.88 & 75.98 & 93.43 & 31.32 & 24.54 & 56.03   \\ \hline
                   & MAE     & 15.96 & 17.48 & 10.81 & 12.47 & 10.87 & 13.52   \\
Seq2Point \cite{bc925f015ff2472d8af9ec4886b91b9e} & SAE     & 10.65 & 8.01  & 5.3   & 10.33 & 8.69  & 8.6     \\
                   & F1 (\%) & 50.92 & 80.32 & 94.88 & 45.41 & 49.11 & 64.13   \\ \hline
                   & MAE     & 14.96 & 16.47 & 12.02 & 10.37 & 9.87  & 12.74   \\
S2SwA \cite{article}     & SAE     & 10.68 & 7.81  & 5.78  & 8.33  & 8.09  & 8.14    \\
                   & F1 (\%) & 53.67 & 79.04 & 94.62 & 47.99 & 45.79 & 64.22   \\ \hline
                   & MAE     & 10.91 & 16.27 & 8.09  & 5.62  & 9.74  & 10.13   \\
SGN \cite{Shin2019SubtaskGN}       & SAE     & 7.86  & 6.61  & 5.03  & 4.32  & 7.14  & 6.2     \\
                   & F1 (\%) & 60.02 & 84.43 & 96.32 & 58.55 & 61.12 & 72.09   \\ \hline
                   & MAE     & 8.88  & 15.25 & 6.20  & 4.90  & 8.35  & 8.71    \\
Proposed Mh-Net CNN     & SAE     & 4.86  & 6.59  & 4.07  & 3.80  & 5.67  & 5.00    \\
                   & F1 (\%) & 63.53 & 87.23 & 97.10 & 61.01 & 63.01 & 74.38   \\ \hline
                   \hline
\end{tabular}}
\end{table}
\section{Conclusion}
This paper has developed a supervised NILM technique using Mh-Net CNN to identify the real-time energy consumption and ToU of individual appliances connected in a building microgrid. The proposed model introduces an attention layer in the CNN architecture which significantly improves model performance by reducing estimation error. Furthermore, the proposed model has been trained with dynamic voltage dataset to improve accuracy in dynamic grid voltage conditions. Appliance sets having varied levels of active and reactive power consumption are experimentally tested to analyze performance under dynamic grid voltage conditions. Results reveal that the active and reactive power profiles vary with variation in supply voltage for majority of the appliances. Further, experimental results highlight that the proposed NILM model is suitable for any appliance combination having wide range of power consumption levels. A comparative analysis of Mh-Net CNN model trained under constant voltage and dynamic voltage conditions is verified through experimental results. The results verify that the performance of the proposed model is improved when subjected to training under dynamic grid voltage conditions. Furthermore, the performance of the proposed model is assessed using well known UK-DALE dataset which is widely adopted in research. The proposed model has been observed to demonstrate superior performance in comparison to the existing techniques in terms of identification accuracy. \vspace{-0.75cm}
\bibliography{Reference}

\end{document}